\documentclass[11pt]{article}
\usepackage[right=1in,left=1in,top=1.1in,bottom=1.1in]{geometry}
\usepackage{hyperref}
\hypersetup{colorlinks, citecolor=blue, filecolor=blue, linkcolor=blue, urlcolor=blue}
\usepackage{graphicx}
\usepackage{url}
\usepackage[round]{natbib}
\usepackage{amsmath,amsthm} 
\usepackage{engord}
\usepackage{float}
\usepackage{subfig}
\usepackage{pdflscape}
\usepackage{booktabs}
\usepackage{pgfplots}
\pgfplotsset{compat=1.14}
\pgfplotsset{every axis label/.append style={font=\tiny}}
\usepackage[labelsep=period]{caption} 

\def\argmin{\mathop{\it argmin}}
\def\argmax{\mathop{\it argmax}}

\usepackage{amssymb} 
\usepackage{multirow} 

\usepackage{xr}

\usepackage{setspace}
\usepackage{sectsty}
\sectionfont{\large}
\subsectionfont{\normalsize}
\subsubsectionfont{\normalsize}

\title{ \vspace*{1cm} \hspace*{-0.5cm}Dynamic Discrete Choice and Inverse Reinforcement Learning:\\
Inferring Preferences and Beliefs From Human Behavior\footnote{John Rust is grateful for support
from his Georgetown University Professorship. This is the authors' accepted manuscript of an article
published in the \emph{Oxford Research Encyclopedia of Economics and Finance} (Oxford University Press,
August 3, 2026). The version of record is available at
\href{https://doi.org/10.1093/9780197851418.003.0990}{https://doi.org/10.1093/9780197851418.003.0990}.
}}

\author{
Pranjal Rawat\thanks{Georgetown University  \href{mailto:pp712@georgetown.edu}{pp712@georgetown.edu}}
\and
John Rust\thanks{Professor Emeritus, Georgetown University
\href{mailto:jr1393@georgetown.edu}{jr1393@georgetown.edu}}}

\date{August 3, 2026}

\begin{document}

\bgroup
\let\footnoterule\relax

\begin{singlespace}
\maketitle

\begin{abstract}
This article surveys two deeply connected literatures that approach the same fundamental problem from different disciplinary traditions: dynamic discrete choice (DDC) in structural econometrics and inverse reinforcement learning (IRL) in machine learning. Both seek to infer the preferences of decision makers from observed sequential behavior, assuming that individuals act to maximize an expected reward function within a dynamic, uncertain environment formalized as a Markov decision process (MDP). Despite independent origins, the two fields have converged on similar mathematical formulations. We show that the ``soft Q-learning'' framework now prevalent in IRL is closely related to DDC models under additive extreme value preference shocks, yielding the same softmax (multinomial logit) choice probabilities and smooth Bellman equations that underpin structural estimation in economics. We compare the estimation and computational methods developed in each field. DDC has emphasized maximum likelihood estimation, conditional choice probability estimators, and policy iteration methods. IRL has developed scalable alternatives, including maximum entropy methods, adversarial approaches, and model-free temporal difference estimators that extend to high-dimensional state spaces using deep neural networks. Model-free IRL estimators that combine temporal difference learning with classical two-step methods from econometrics represent a promising direction for bridging the two literatures. Both fields confront shared foundational challenges: the identification problem, whereby multiple reward functions can rationalize the same observed behavior, and the curse of dimensionality in solving the underlying MDP. Each literature addresses these challenges under different assumptions and with different objectives, DDC prioritizing counterfactual policy evaluation and IRL prioritizing training agents from expert demonstrations. We believe that cross-fertilization offers substantial opportunities for methodological progress in both fields.

\end{abstract}
\end{singlespace}
\thispagestyle{empty}
\clearpage
\egroup
\setcounter{page}{1}

\section{Introduction\label{sec:introduction}}

This article compares {\it dynamic discrete choice\/} (DDC), a subfield of structural econometrics (SE),
to {\it inverse reinforcement learning\/} (IRL), a subfield of machine learning (ML). 
The goal of both is to infer preferences of individuals from observations of their choices over time 
assuming they are {\it rational,\/} i.e. that their choices over time maximize an expected reward (or ``utility'') function based on 
beliefs of how their actions affect uncertain current and future outcomes and payoffs. 
Individual  choices are affected by stochastic shocks that make their behavior appear probabilistic
from the standpoint of outside observers. DDC and IRL share a common framework, Markovian decision processes (MDP)
that can be viewed as dynamic versions of {\it random utility models\/} in economics, allowing
inference of preferences by methods such as
maximum likelihood, see \citet{mcfadden:1973} and \citet{rust:1987}. DDC and IRL differ primarily in the methods used to
solve the MDP, conduct inference, as well as the overall rationale for inferring preferences.

SE focuses on  testing and evaluating economic theories and making {\it counterfactual predictions\/}
of the effect of changes in the environment and
economic policies on economic behavior, outcomes, and welfare.\footnote{\footnotesize In economics ``policy'' refers to government interventions (tax rates, regulations, etc.), whereas
in RL/MDP it refers to a {\it decision rule,\/} i.e. a mapping of states into a probability distribution over actions.} 
SE differs from {\it reduced-form econometrics\/} that focus on summarizing behavior without attempting to infer 
preferences and thus without requiring the rationality assumption, analogous to the corresponding distinction
 between IRL and {\it imitation learning\/} (IL) in the ML literature.\footnote{\footnotesize
Other names for this include {\it behavioral cloning,\/} {\it learning from demonstration\/} and {\it apprenticeship learning.\/}
These are types of {\it supervised learning\/} in the ML literature, where an 
individual's decision rule can be learned from sufficient data on pairs of actions and states, $(a,s)$, without
any assumption that actions taken are ``optimal'',  see \citet{hussein:2018}.}
Reduced-form models can be used to make
some types of counterfactual predictions when there is sufficient
historical variation in policies to allow them to be treated as covariates in the model.
By uncovering preferences, SE models allow us to assess the welfare impacts of policy changes that are not
possible using reduced-form models and enable 
counterfactual predictions of policy changes that have no close historical antecedent.

For example, Denmark has long had one of the world's highest tax rates on new cars, but with little historical variation 
that could be used by reduced form approaches to predict how reductions in this tax would affect the Danish economy.
\citet{GIMRS:2022} structurally estimated an equilibrium model of ownership and trading of cars to recover consumer preferences
under the high tax {\it status quo\/} regime, and used it to compute counterfactual equilibria under a variety of alternative tax rates.
They showed that by reducing the tax on new cars and increasing the gas tax, Denmark can: 1) improve the welfare of most of its citizens,
2) increase total tax revenue from cars, and 3) reduce total $\mbox{CO}_2$ pollution.
Credible structural models are a faster, cheaper way of predicting the effects of policy changes by 
``crash testing'' them with computer simulations before actually implementing them in reality.

IRL, introduced by \citet{ng:2000}, builds on {\it reinforcement learning\/} (RL) \citep{BartoSutton:1998}. Its goal is to develop intelligent agents that can perform complex robotic and intellectual tasks without
the need for explicit programming or an explicit reward function to direct behavior. The success of ML 
is largely due to supervised learning algorithms (such as neural networks) to predict outcomes using labeled data. 
However this is data-intensive and does not naturally handle sequential decision making 
where actions have long-term consequences. RL overcomes this by training agents to maximize a reward through trial-and-error,
via repeated dynamic interactions with the environment. 
When the reward function is known, RL can  approximate optimal behavior without labeled examples. Chess is an example with a known reward (1 for a win, -1 for a loss and 0 for a draw). RL has been able to learn superhuman performance via {\it self-play,\/} i.e. repeatedly playing the algorithm against itself resulting in performance that exceeds the best human and computerized chess players, see e.g. \citet{Silver:2018}.

Unfortunately, it is often unclear what reward function to use to program tasks in AI and robotic applications using RL.
For example, \citet{christiano:2017} note that RL provides no guidance on the appropriate reward function to 
``train a robot to clean a table or scramble an egg.''  
IRL solves this by learning a reward function from limited data on human or expert demonstrations. Using the inferred reward function, 
these algorithms can be  trained more rapidly and cheaply via RL using computer simulations that result in
 intelligent behavior consistent with human preferences in situations outside of those in a limited demonstration
data set.
For example, Google Maps trip data \citep{Barnes:2024} revealed travelers' route preferences reflecting trade-offs between traffic, distance, hills, safety, and scenery. They massively scaled IRL to estimate a route preference function (360M parameters, 110M trips) enabling RL to generate recommendations achieving 16-24\% improvement in global route quality, and to the best of our knowledge, represents the largest published application of IRL in a real-world setting to date.

Section 2 reviews Dynamic Programming (DP) and Markovian Decision Processes (MDP), the common framework
used in the DDC and IRL literatures, and reinforcement learning (RL). Section 3 reviews the DDC literature
and section 4 the IRL literature. Section 5 concludes, comparing/contrasting
the literatures and the key challenges and limitations they both face.

\section{Background on DP and MDPs\label{sec:mdp}}

DDC and IRL apply  the recursive mathematical characterization of optimal sequential decision
making under uncertainty known as  {\it dynamic programming\/} (DP), \citet{Bellman:1957}.\footnote{\footnotesize See
\citet{bertsekas:2017} for an excellent introduction to the DP literature.} We focus 
on stationary, infinite horizon Markovian Decision Problems (MDP), a subclass of DPs introduced by Bellman and \citet{Howard:1960} that 
are the dominant paradigm in SE and RL.
MDPs are defined by  1) a state space $S$, 2) an action space $A$, and  3) the objects  $\{r,p,\beta\}$, where $r(s,a)$ is the {\it reward or utility function\/}
providing the payoff to a decision maker (DM), $p(s'|s,a)$ is a {\it transition density\/} for next period's state when the state is $s$ and action is $a$
and $\beta \in (0,1)$ is a discount factor.  To simplify exposition and avoid technicalities, we assume
that  $S$ and $A$ are finite sets with $|S|$ and $|A|$ elements, respectively.
A policy or {\it decision rule\/} $\pi$ is a conditional probability of choosing 
a feasible action $a \in A(s)$ for all $s \in S$. It is called a {\it mixed strategy\/} in game theory  to 
distinguish it from a {\it pure strategy,\/} i.e. a deterministic function $a=\delta(s)$ that
chooses a single optimal action with probability 1. Since mixed strategies include pure strategies as a special case, 
we can express the DM's objective as the choice of a policy
$\pi$ that  maximizes expected discounted rewards given by
\begin{equation}
       V(s)= \max_\pi  V_\pi(s),  \quad \mbox{where} \;  V_\pi(s) = E_\pi \left\{\sum_{t=0}^\infty \beta^t r(s_t,a_t)| s_0=s\right\},
\label{eq:vdef}
\end{equation}
where $E_\pi$ denotes the expectation operator over the Markov process $\{s_t,a_t\}$ implied by the transition probability $p$ and policy $\pi$.
The optimal value function can be defined recursively via {\it Bellman's equation\/}
\begin{equation}
            V(s) = \max_{a \in A(s)}\left[ r(s,a)+ \beta \sum_{s' \in S} V(s')p(s'|s,a)\right] \equiv \Gamma(V)(s).
\label{eq:bellman}
\end{equation}
$V$ is the fixed point $V=\Gamma(V)$ to an operator $\Gamma(V)$ known as the ``Bellman operator'' 
which can be shown to be a {\it contraction mapping\/} 
(i.e. for any vectors $V,W \in R^{|S|}$ we have $\|\Gamma(V)-\Gamma(W)\| \le \beta \|V-W\|$ where $\beta$ is less than 1
and $\|\cdot\|$ is the Euclidean norm). This implies that a  unique solution $V$ to Bellman's equation exists, 
and the optimal policy $\pi$ takes the form of a pure strategy, i.e. a deterministic function $a=\delta^*(s)$
that can be recovered from $V$ by
\begin{equation}
       \delta^*(s) = \argmax_{a \in A(s)}\left[ r(s,a)+\beta \sum_{s' \in S} V(s')p(s'|s,a)\right],
\label{eq:bellman_policy}
\end{equation}
since generically a unique action $a^*=\delta^*(s)$ attains the max on the right hand side of (\ref{eq:bellman_policy}).
Thus, a mixed strategy $\pi$ cannot be optimal in general. This appears to challenge the relevance of the MDP
framework since most strategies we estimate/learn empirically will be mixed rather than pure. 
The DDC literature resolves this by assuming that  
we only observe a subset of the full state vector the individual observes. This allows an optimal
pure strategy to appear to us as a mixed strategy given our inability to observe the agent's full information set. 
It also facilitates statistical inference of the MDP as we show in Section 3.
We now summarize the key methods for solving for  $V$ and $\delta^*$.

\subsection{Successive Approximations}

Since a contraction mapping $\Gamma$ has a unique fixed point $V=\Gamma(V)$, the contraction property guarantees that 
the sequence $\{V_k\}$ given by $V_k=\Gamma(V_{k-1})$, for $k=1,2,\ldots$ starting from
any initial starting guess $V_0$ converges to the fixed point $V=\Gamma(V)$ at a geometric rate in $\beta$,
which is very slow when $\beta$ is close to 1. 

\subsection{Policy Iteration}
\label{section:pi}

The \citet{Howard:1960} {\it policy iteration\/} (PI) algorithm is a faster method for solving MDPs that 
alternates between {\it policy valuation\/} and {\it policy improvement\/} steps. At iteration $k$ the PI algorithm has a candidate
policy $\delta_k$. The policy valuation step calculates the value of $\delta_k$,  $V_{\delta_k}$, by solving the linear system
\begin{equation}
     V_{\delta_k} = r_{\delta_k} + \beta E_{\delta_k} V_{\delta_k},
\label{eq:policy_valuation}
\end{equation}
where $r_{\delta_k}$ is the vector whose $s^{th}$ element is $r(s,\delta_k(s))$ and $E_{\delta_k}$ is the $|S| \times |S|$ matrix whose
element in row $s$ and column $s'$ is given by $E_{\delta_k}(s,s')=p(s'|s,\delta_k(s))$. For any feasible deterministic policy $\delta$
the matrix $E_{\delta}$ is a Markov transition probability matrix with matrix norm $\|E_{\delta}\|=1$, which implies that equation (\ref{eq:policy_valuation})
has a unique solution given by $V_{\delta_k}=[I-\beta E_{\delta_k}]^{-1} r_{\delta_k}$. Given the value $V_{\delta_k}$ the  policy improvement step
calculates a new decision rule $\delta_{k+1}$ given by the right hand side of equation (\ref{eq:bellman_policy}) except with 
$V_{\delta_k}$ substituted for $V$. 
If $\delta_{k+1}=\delta_k$ then policy iteration has converged and it is not hard to show that 
$V_{\delta_k}=V$, the unique solution to Bellman's equation (\ref{eq:bellman}).  

\subsection{Randomization, function approximation, and the Curse of Dimensionality}
\label{section:curse}

Inference in DDC and IRL requires repeated solutions of the MDP to find a reward function whose implied optimal behavior
best approximates observed behavior according to different metrics of goodness of fit, such as maximizing a likelihood 
function.  It is already challenging to solve a single MDP when $|S|$ or $|A|$ is large since each SA 
step requires $O(|A||S|^2)$ arithmetic operations to compute the expectation and maximization operators 
in equation (\ref{eq:bellman}) for all $s \in S$. Though PI requires only a few iterations to converge and thus
can be faster for smaller problems when $\beta$ is near 1, the downside is the $O(|S|^3)$ operations
required to solve the linear system (\ref{eq:policy_valuation}) in each policy valuation step.
In the absence of sparsity or other ``special structure'' alternative methods are required when $|S|$ or $|A|$ is large.

Bellman called this  the {\it curse of dimensionality\/} because it arises 
when  discretization and interpolation are used to approximate solutions to MDPs with multiple continuous state variables. 
Approximating a continuous MDP with $d$ state variables using a finite grid of $N$ points in each dimension 
results in an approximate finite MDP with $|S|=O(N^d)$ states which grow exponentially in $d$. The curse of dimensionality is unavoidable \citep{CT:1989}:  MDPs with $d_s$ continuous state and $d_c$ continuous choice variables
have a worst-case complexity (minimum number of arithmetic operations to compute an $\varepsilon$-approximation to $V$) 
that is bounded below by a function that grows at rate  $1/[(1-\beta)\varepsilon)]^{2d_s+d_c}$, a  lower bound that holds for {\it any possible 
algorithm.\/}

It may be possible to break the curse of dimensionality for subclasses of MDPs with special structure, such as  DDCs 
where the choice set $A$ is finite. When $|A|$ is small, most of the work to solve the MDP is spent on 1) numerical integration to
compute conditional expectations $EV(s,a)=\int V(s')p(s'|s,a)$, and 2) interpolation/approximation needed to find $V(s')$ for $s' \in S$.
While randomization techniques such as Monte Carlo integration have been proposed to mitigate the curse of dimensionality \citep{rust:1997}, their general applicability remains debated \citep{bray:2022}.

Another way to break the curse of dimensionality is to approximate $V$ by a function $V_\phi$  
that depends on a small number of unknown parameters $\phi$ for which $V_\phi$ approximately
solves the Bellman equation. For example $V_\phi$ could be the linear combination
$V_\phi(s)=\sum_{k=1}^K \phi_k \rho_k(s)$ where $\{\rho_k\}$ are ``basis functions'' and $\phi=(\phi_1,\ldots,\phi_K)$ are
coefficients, or $V_\phi$ could be a nonlinear function of $\phi$ such as a neural network.
Neural networks are universal approximators, which might help break the curse of dimensionality of approximating $V$, 
at least for certain subclasses of sufficiently smooth functions as first demonstrated by \citet{Barron1993}.
Let $V_{\hat\phi}$ denote 
a best approximation to $V$, where $\hat\phi$  
minimizes the mean squared ``Bellman residuals'' $\mbox{BE}(\phi)$
 over some grid of $N$ points $S_N=(s_1,\ldots,s_N)$ in $S$
\begin{equation}
      \hat\phi = \argmin_{\phi} \mbox{BE}(\phi)=E_{s \sim S_N}\left\{\left[V_\phi(s)- \Gamma(V_\phi)(s)\right]^2\right\},
\label{eq:nlls}
\end{equation}
where $E_{s \sim S_N}\{h(s)\}=\frac{1}{N} \sum_{i=1}^N h(s_i)$ denotes the sample average over the $N$ points in $S_N$.
Neural net approximation breaks the curse of dimensionality by allowing us to approximate $V$ with a small number
of network weights/parameters $\phi$.  Randomization also helps break the curse of dimensionality via
Monte Carlo integration instead of deterministic quadrature to approximate the  
conditional expectations $EV(s,a)$ entering the Bellman operator
$\Gamma(V_\phi)(s)$, transforming the solution of the Bellman equation for $V$ into a nonlinear regression problem (\ref{eq:nlls}).

Using the Triangle inequality and the contraction property of $\Gamma$, we can derive the error bound
$\|V-V_\phi\|_\infty \le \| V_\phi-\Gamma(V_\phi)\|_\infty/(1-\beta)$.\footnote{\footnotesize Here $\|f\|_\infty$
denotes the supremum norm of the function $f$, $\|f\|_\infty=\sup_{s \in S}\|f(s)\|$.} Thus, using a richer class of functions 
$\mathcal{F}_K=\{V_\phi|\phi \in R^K\}$ (i.e. a larger $K$) we can make the Bellman error
$\|V_\phi-\Gamma(V_\phi)\|_\infty$ smaller and better approximate the true solution to the Bellman
equation $V=\Gamma(V)$. In general, we require both $N$ and $K$ to increase to infinity at appropriate rates
to manage a classic bias-variance tradeoff to ensure that the nonlinear least squares estimator $V_{\hat\phi}$ (\ref{eq:nlls})
is a consistent estimator of the true $V$. Even when we can approximate $V$ well with relatively few parameters
$K$, there may still be a curse of dimensionality associated with finding a global minimizer $\hat\phi$.\footnote{\footnotesize 
Though stochastic optimization methods such as stochastic gradient descent can be helpful in avoiding getting stuck
at local minima, randomization does not break the curse of dimensionality of continuous non-convex optimization \citep{yn:1976}.}
Further, when $\beta$ is close to 1 $V_{\hat\phi}$ is not guaranteed to be close to the
true $V$. For example, if we start with 
the true solution $V=\Gamma(V)$ and perturb it by adding a constant $C$, the perturbed solution has a Bellman error of
$(1-\beta)C$ which can be small even when $C$ is large if $\beta$ is close to 1.
Fortunately, DDCs do not require
precise approximation of the level $V$ to accurately approximate the optimal policy in  (\ref{eq:bellman_policy}) since shifting $V$
by a constant leaves $\delta^*$ unaffected.\footnote{\footnotesize \citet{fujimoto22a} show that Bellman error is a poor proxy for value error in offline RL. 
\citet{Nguyen:2025} shows that in DDC problems, we obtain
more accurate solutions for $\delta^*$ by minimizing the sum of squared {\it anchored\/} Bellman errors, 
$V_\phi(s_j)-V_\phi(s_1)-[\Gamma(V_\phi)(s_j)-\Gamma(V_\phi)(s_1)]$ that difference out
parallel shifts in $V_\phi$, allowing the minimizer to more accurately approximate the {\it shape\/} of $V$ by ignoring its level, 
which is a version of ``relative value iteration'' proposed by \citet{puterman:1994} for solving MDPs under the long run average criterion.}

\subsection{Reinforcement Learning}
\label{section:rl}

These are stochastic iterative algorithms for approximating the value function $V$ and the optimal
policy $\pi$ inspired by models of trial and error learning by animals and 
humans.\footnote{\footnotesize  See \citet{BartoSutton:2020} and \citet{bertsekas:2025}
for excellent introductions to this literature.}
One of the first RL methods was 
{\it temporal difference learning\/} (TD) \citep{sutton:1988} which can be regarded as a type of asynchronous 
{\it stochastic approximation\/} \citep{robbinsmunro:1951} for solving the linear system (\ref{eq:policy_valuation}) for
$V_\delta$ in the policy valuation step of PI. TD is a special case of  
{\it Q-learning\/} \citep{Watkins:1989} where $Q$ is the {\it choice-specific value function\/} implied by the  Bellman equation
\begin{equation}
 \label{eq:q1}  
      Q(s,a)  =  r(s,a) + \beta \sum_{s'} V(s')p(s'|s,a)
  = r(s,a)+\beta \sum_{s'} \max_{a' \in A(s')} Q(s',a')p(s'|s,a) \equiv \Lambda(Q)(s,a).
\end{equation}
$Q$-learning can be viewed as a stochastic version of successive approximations 
$\{Q_t\}$ for finding the fixed point $Q=\Lambda(Q)$ via the updating rule
\begin{equation}
   Q_{t+1}(s_t,a_t)= Q_t(s_t,a_t) + \alpha_t \left[ r(s_t,a_t)+\beta \max_{a \in A(s_t)}[Q_t(\tilde s_{t+1},a)] - Q_t(s_t,a_t) \right],
\label{eq:qlearning}
\end{equation}
where $(s_t,a_t)$ is the state and action taken at  $t$, and $\tilde s_{t+1}$ is a random draw from the transition probability
$p(s'|s_t,a_t)$. \citet{Tsitsiklis:1994}
showed that if the step size sequence $\{\alpha_t\}$ converges to 0 at a sufficiently slow rate, the sequence $\{Q_t\}$ converges with probability 1 
to the unique $Q$ that solves (\ref{eq:q1}).\footnote{\footnotesize  \label{fn:step_size}
The conditions on the step size are 1) $\sum_t \alpha_t = \infty$
and 2) $\sum_t \alpha_t^2 < \infty$. Note $\alpha_t=1/t$ satisfies these conditions.}
Equation (\ref{eq:qlearning}) does not specify how  $a_t$ is chosen: optimality requires a {\it greedy policy\/}
$a_t = \argmax_{a \in A(s_t)} Q_t(s_t,a)$, but a key requirement for the convergence of Q-learning is 
that all pairs $(s,a)$ are sampled infinitely often. This presents an exploration/exploitation 
dilemma that implies that suboptimal actions must
be taken infinitely often in order for $Q$ learning to converge, but this
cannot occur under a greedy policy which is generically a pure strategy.  
One solution is to use an {\it $\varepsilon$-greedy policy\/} that takes the greedy action $a_t$
 with probability $1-\varepsilon$, and with probability $\varepsilon$ a randomly selected 
suboptimal action. However unless $\varepsilon$ converges
to $0$ with $t$, the policy we use to learn $Q$ is necessarily suboptimal and different from the optimal policy  we are trying to learn.
Thus Q-learning is an example of what \citet{BartoSutton:2020} refer  to as  {\it off-policy learning.\/}

\citet{softqlearning:2017} resolved the exploration/exploitation tradeoff  by 
 building on the {\it maximum entropy reinforcement learning\/} framework of \citep{ziebart:2008},
resulting in an algorithm they call {\it soft-Q learning.\/} 
They used Q-learning to solve a modified  MDP where the reward
 includes an entropy term that forces the optimal policy to be mixed rather than pure.
This ensures that Q-learning results in continuous exploration with all feasible actions being
chosen infinitely often, and they showed
it converges with probability 1 to the optimal mixed strategy $\pi$ given by the classic logit
or softmax formula that we will derive in equation (\ref{eq:ccp}) of section~\ref{sec:ddc}. 
The entropy of $\pi$ is $H(\pi(s))=-\sum_{a \in A(s)}\pi(a|s)\log(\pi(a|s))$. Adding it to the
reward results in the  MDP
\begin{equation}
       V(s)= \max_\pi  V_\pi(s),  \quad \mbox{where} \;  V_\pi(s) = E_\pi \left\{\sum_{t=0}^\infty \beta^t[r(s_t,a_t)+\sigma H(\pi(s_t))]| s_0=s\right\}.
\label{eq:vdefentropy}
\end{equation}
When $\sigma=0$ the objective (\ref{eq:vdefentropy}) reduces to the original MDP objective
(\ref{eq:vdef}) where $\pi$ is the pure strategy (\ref{eq:bellman_policy}). 
As $\sigma \to \infty$ the optimal policy for (\ref{eq:vdefentropy}) converges to a uniform distribution,
the maximum entropy solution. For $0 < \sigma < \infty$ the optimal
policy $\pi$ takes the form of a multinomial logit or ``softmax'' function (\ref{eq:ccp})
and the optimal value function $V$  in equation  
(\ref{eq:vdefentropy}) satisfies a ``soft Bellman equation'' identical
to the ``smoothed Bellman equation'' derived by \citet{rust:1988} for a ``partially observed'' MDP 
in equations  (\ref{eq:smoothV}) and (\ref{eq:smoothQ}) of section~\ref{sec:ddc}. In DDC models, individuals
use optimal pure strategies and the
entropy term $\sigma H(\pi(s))$ equals the expectation of components of rewards  that individuals
observe but we  do not observe. In contrast, the IRL literature uses entropy 
to enforce a direct preference for using mixed rather than 
pure strategies.\footnote{\footnotesize From an economic standpoint, it is not clear why individuals
should care about the entropy of a policy $\pi$: instead the random utility framework posits that
individuals care about the policy they use only to the extent that it directly maximizes their rewards.  
\citet{matejka:2015} provide an interpretation of the random utility model as a Bayesian decision problem
that involves a first stage choice of how much information to acquire about a finite set of alternative
choice options, where entropy captures the cost of information. They show that under an optimal
information acquisition strategy, the second stage choice probabilities take the multinomial or
``softmax'' form (\ref{eq:ccp}) derived in section~\ref{sec:ddc}. However this is a one shot model
of information acquisition: extending it to  DDC models involving sequential 
choices requires a Bayesian MDP formulation that includes
a component of the state variable that captures the posterior distribution of the individual's beliefs about the rewards
from various choices which is gradually updated over time, reflecting ``learning by doing.'' A classic example is the
``multi-armed bandit problem'' \citet{gittins:1974}.}

We say that RL is \textit{model-free} if it does not require explicit specification or knowledge of $p(s'|s,a)$.
Otherwise it is \textit{model-based}.  Q-learning can be  {\it model-free,\/}  unlike SA or PI, to the
extent that it does not require a specification for the
transition probability $p(s'|s,a)$ and avoids numerical integration to update $V$ and $Q$. This is possible
when ``the environment'' generates new states  rather than a computer simulation based on an explicit 
mathematical model for $p(s'|s,a)$. In  Watkins'  example
of an animal learning to adapt and thrive, we do not need to know $p$ since the environment ``draws'' the successor states
$s_{t+1}$ needed to implement the update in $Q_t$ in equation (\ref{eq:qlearning}). When trained in this manner Q-learning is indeed model-free
and so are other RL algorithms that interact directly with the environment (or use historical transitions for training/estimation as we discuss
in section~\ref{sec:irl}).

In many situations Q-learning and related RL methods produce policies that
are too noisy and converge too slowly to be useful in 
real applications, especially when the number of states is large.  As \citet[p.~1]{JiangXie:2024} note, their
``decisions can lead to undesirable outcomes, especially at the early stages of learning when the algorithm has little knowledge of the environment. This is not a problem when the environment is a simulator, but can lead to serious consequences when people are part of the environment.''
As a result RL training is increasingly done in simulation environments with
sufficient computer power to rapidly conduct many thousands or millions of training iterations to produce a much better solution than could be achieved
from limited real-world experience.  The ability to simulate from the true $p(s'|s,a)$  (or a good approximation to it)
for a large number of iterations is  key to producing good solutions using RL. For this reason most of the successful recent applications
of RL are  model-based in the sense that they rely on a model $p(s'|s,a)$ to simulate outcomes in offline
training even while using model-free RL algorithms to train the $Q$ values.\footnote{\footnotesize The use of Q-learning
to train chess strategies is an interesting intermediate case. Although the functional form of $p(s'|s,a)$
for chess is unknown since this transition probability depends on the action of the opponent, Q-learning can be classified as 
model-free since it relies on the environment to produce the next state $s_{t+1}$ directly
via self-play, without explicitly modeling or using the game's rules to calculate $p(s'|s,a)$.
In contrast, an algorithm like AlphaZero that uses Monte Carlo tree search explicitly uses the
game rules for planning and can be classified as model-based.}  

Tabular (i.e. finite state/action) Q-learning, (\ref{eq:qlearning}), is far too slow for MDPs with large state or action spaces. 
A scalable solution to this problem is to use function approximation methods 
such as Deep Q-Networks (DQN), which are  deep neural networks with weights $\phi$ 
that parameterize the Q-function as $Q_\phi$ similar to the way $V$ is approximated by $V_\phi$ in (\ref{eq:nlls}).
Instead of iteratively updating Q-values, as in (\ref{eq:qlearning}), 
the goal is to find $\hat\phi$ that minimizes the mean squared Bellman error
\begin{equation} 
   \mbox{BE}(\phi) = E_{(s,a) \sim \mathcal{D}} \left\{\left[\Lambda(Q_\phi)(s,a)-Q_\phi(s,a)\right]^2\right\}
=E_{(s,a) \sim \mathcal{D}}\left\{\left[ r(s,a)+\beta EV_\phi(s,a) -Q_\phi(s,a)\right]^2\right\},
\label{eq:dqnideal}
\end{equation}
where $\mathcal{D}=\{(s_i,a_i)|i=1,\ldots,N\}$ is a deterministic or random grid of $(s,a)$ pairs and $EV_\phi(s,a)$ is the
``emax'' function $\sum_{s'} \max_{a' \in A(s')} Q_\phi(s',a')p(s'|s,a)$.
Minimizing $\mbox{BE}(\phi)$ can be a difficult, high dimensional
optimization problem and we cannot even calculate it if the transition probability
$p$ is unknown.  There is also a danger that overfitting $Q$ can produce solutions with
low Bellman error $\mbox{BE}(\hat\phi)$ but suboptimal implied policies $\pi_{\hat\phi}$ \citep{fu2019diagnosing}.
However there are ways to deal with these challenges, and minimizing $\mbox{BE}(\phi)$ is the most common
objective when using DQN and other parametric methods to find approximate solutions to large scale MDPs.
Some methods instead target the mean squared projected Bellman error, recognizing that the range of the
Bellman operator need not lie in the approximating class.

An important goal in RL  is to try to avoid numerical integration with respect to  $p(s'|s,a)$ 
entering $\mbox{BE}(\phi)$ to obtain model-free learning similar to tabular Q-learning, (\ref{eq:qlearning}). This
suggests modifying the $\mbox{BE}(\phi)$ criterion to choose to $\hat\phi$ to minimize mean squared {\it temporal difference error\/}
\begin{equation}
      \mbox{TD}(\phi) = \mathbb{E}_{(s,a,s')\sim\mathcal{D}} \left\{ \left[r(s,a) + \beta \max_{a' \in A(s')} Q_{\phi}(s', a') - Q_\phi(s, a) \right]^2 \right\}.
\label{eq:td_loss}
\end{equation}
Unfortunately, it is well known that the $\hat\phi$ that minimizes $\mbox{TD}(\phi)$ 
differs from the value that minimizes $\mbox{BE}(\phi)$, since it is easy to see that
\begin{equation}
         \mbox{TD}(\phi) = \mbox{BE}(\phi) + \beta^2 \mathbb{E}_{(s,a,s')\sim\mathcal{D}} \left\{ \left[ \max_{a' \in A(s')} Q_\phi(s',a')
- EV_{\phi}(s,a)\right]^2\right\},
\label{eq:BETDdiff}
\end{equation}
a classic ``bias-variance'' decomposition,  where the second term on the right hand side of (\ref{eq:BETDdiff}), $\mbox{VQ}(\phi)$, is the
variance of the TD error.  

We now describe three main  modifications to the TD criterion (\ref{eq:BETDdiff}) that are used in current applications
of RL to achieve the benefits of model-free
simulation and training while still consistently estimating  the true $Q$ and $V$ in  (\ref{eq:q1}).
The first,  proposed by \citet{antos:2008}, is to parametrize the emax function $EV$ using separate parameters $\psi$
as  $EV_\psi(s,a)$ and use (\ref{eq:BETDdiff}) to write the Bellman error as $\mbox{BE}(\phi,\psi)=\mbox{TD}(\phi)-\beta^2 \mbox{VQ}(\phi,\psi)$,
and solve the minimax problem 
$(\hat\phi,\hat\psi)=\argmin_\phi\argmax_\psi \mbox{BE}(\phi,\psi)$. They showed that 
this modification results in a consistent estimator of $Q$ and $V$, but with the downside that the resulting minimax problem may result
``in considerable additional computational burden''. 
A second approach is to modify $\mbox{TD}$ to include
a moving  ``target function'' for $Q$, resulting in an iterative approach to minimizing BE called
{\it fitted value iteration\/} (FVI) \citet{munos:2008}.  Suppose we have some estimate of $\phi$  at
iteration $t-1$, $\hat\phi_{t-1}$. FVI results in an updated estimate $\hat\phi_t=\argmin_\phi\mbox{TD}(\phi,\hat\phi_{t-1})$
where the modified TD criterion using the target function $Q_{\hat\phi_{t-1}}$ is 
\begin{equation}
 \mbox{TD}(\phi,\hat\phi_{t-1}) = \mathbb{E}_{(s,a,s')\sim\mathcal{D}} \left\{ \left[r(s,a) +
 \beta \max_{a' \in A(s')} Q_{\hat\phi_{t-1}}(s', a') - Q_\phi(s, a) \right]^2 \right\}.
\label{eq:dqn_target}
\end{equation}
The target function ``trick'' converts the original TD objective (\ref{eq:td_loss}) into a standard linear or nonlinear regression, 
and it is easy to see that by doing this, the variance term in (\ref{eq:BETDdiff}) no longer depends on $\phi$, so when 
$Q_{\hat\phi_{t-1}}$ is close to the true $Q$, $\mbox{BE}(\phi)$ and $\mbox{TD}(\phi,\hat\phi_{t-1})$ are both minimized
by the same value $\hat\phi_t$, and the implied $Q_{\hat\phi_t}$ will be close to the true $Q$. 
The third approach is referred to as a {\it semi-gradient method\/} (see section 9.3 of
\citet{BartoSutton:2020}) since it equals the $\hat\phi$ that solves 
(or approximately solves) the $K$ ``moment conditions''
\begin{equation}
           0 = E_{(s,a,s') \sim \mathcal{D}} \left\{ \left[r(s,a)+\beta  \max_{a' \in A(s')} Q_{\phi}(s', a')
 -Q_\phi(s,a)\right]\nabla_\phi Q_\phi(s,a)
\right\},
\label{eq:semigradient_objective}
\end{equation}
where $\nabla_\phi Q_\phi(s,a)$ is the gradient of $Q_\phi$ with respect to $\phi$.\footnote{\footnotesize
It is referred to as a ``semigradient'' since the expression on the right hand side
of (\ref{eq:semigradient_objective}) is not a full gradient of  $\mbox{TD}(\phi)$
with respect to $\phi$ in (\ref{eq:td_loss}), but only with respect to the last term, $Q_\phi(s,a)$.}
If $Q=Q_{\phi^*}$, then it is easy
to see that for each $(s,a)$ the conditional expectation of $r(s,a)+\beta  \max_{a' \in A(s')} Q_{\phi}(s', a')-Q_\phi(s,a)=0$,
and hence the moment conditions defining the semi-gradient estimator  hold at $\phi=\phi^*$.
In the case of a linear approximator, if $Q(s,a)=\vec{\rho}(s,a)^{\top}\phi^*$ where $\vec{\rho}=(\rho_1,\ldots,\rho_K)$, then 
the $\phi^*$ that solves (\ref{eq:semigradient_objective}) is given by the {\it least squares TD\/} formula in section 9.8 of \citet{BartoSutton:2020},
\begin{equation}
       \phi^*= \left[E_{(s,a,s') \sim \mathcal{D}}\left\{ \vec{\rho}(s,a)^{\top}[\vec{\rho}(s,a)-\beta \max_{a' \in A(s')}\vec{\rho}(s',a')]\right\}\right]^{-1}
 E_{(s,a) \sim \mathcal{D}}\left\{\vec{\rho}(s,a)^{\top} r(s,a)\right\}.
\label{eq:lstd}
\end{equation}
Training of parameters in RL and ML models is customarily done sequentially and incrementally using 
{\it stochastic gradient descent\/} (SGD) where the parameters $\phi$ are updated in lockstep as each
new observation $(s_{t+1},s_t,a_t)$ arrives in  a simulated trajectory rather than only periodically after
the estimation objective and its gradient is evaluated for the full sample (or batch) of simulated outcomes $\mathcal{D}$. 
SGD can be viewed as the application of stochastic approximation to find a parameter $\phi$ that sets the
expected value of a system of equations to zero. In an optimization context such as RL, this system might be
the expected gradient of the objective function we are trying to minimize. 
For example consider the linear semigradient estimator for $\phi$ which solves (\ref{eq:semigradient_objective}).
Instead of directly solving it with, say, Newton's method, SGD solves it via the iterations
\begin{equation}
           \phi_{t+1}=\phi_t + \alpha_t\left[r(s_t,a_t)+\beta \max_{a' \in A(s_{t+1})} Q_{\phi_t}(s_{t+1},a')-Q_{\phi_t}(s_t,a_t)\right]
\nabla_\phi Q_{\phi_t}(s_t,a_t),
\label{eq:sgd_linearsemigradient}
\end{equation}
where $\alpha_t > 0$ is a step size parameter that satisfies the same conditions
as for tabular Q-learning in footnote~\ref{fn:step_size}. \citet{melo:2008} showed that if we approximate $Q$
as a linear combination of $K$ features $\vec{\rho}(s,a)$, $Q_\phi(s,a)=\vec{\rho}(s,a)^{\top}\phi$, then iterations of the form 
(\ref{eq:sgd_linearsemigradient}) converge with probability 1  to a parameter $\phi^* \in R^K$ satisfying
$Q_{\phi^*}=P_K(\Lambda(Q_{\phi^*}))$ where $P_K$ is the orthogonal projection operator onto the linear subspace 
spanned by the $K$ functions $\vec{\rho}(s,a)$. If $Q$ is in this subspace, then $Q=Q_{\phi^*}$, 
$P_K\Lambda=\Lambda$, and  $\phi^*$ is given by (\ref{eq:lstd}).

Counterexamples due to \citet{Baird:1995} and \citet{TsitsiklisVanRoy:1997} show that linear semigradient TD learning methods
such as (\ref{eq:sgd_linearsemigradient}) can diverge unless special care is taken. \citet{BartoSutton:2020} refer to the problem
as the  {\it deadly triad\/}  that can arise when off-policy training is combined with function approximation
and a  ``bootstrapping target'' (i.e. the 
term  $\max_{a' \in A(s')} Q_{\phi_t}(s_{t+1},a')$ in (\ref{eq:sgd_linearsemigradient})) that
changes too rapidly over iterations $t$. \citet{melo:2008} prove convergence under  
a condition on the moment matrix $E_\pi\{\vec{\rho}(s,a)^{\top}\vec{\rho}(s,a)\}$ of the basis functions
$\vec{\rho}(s,a)$ that guarantees that ``the trajectories of the algorithm to closely follow those of an associated
ODE with a globally asymptotically stable equilibrium point.'' (p. 666). 

However the main approach to ensuring stability of iterations with nonlinear approximators such as DQN is to use a ``target network''
that changes more slowly across iterations $t$.  \citet{munos:2008} provided sufficient conditions for convergence in probability 
of FVI iterations $\hat\phi_t =\argmin_\phi \mbox{TD}(\phi,\hat\phi_{t-1})$ where $\hat\phi_{t-1}$
constitute the target network parameters. However their analysis assumes  $\hat\phi_t$ is computed
in ``batch mode'' (i.e. using the entire simulated dataset $\mathcal{D}$) rather than updated recursively using
SGD.  \citet[p.~12621]{zhang:2019} showed that SGD is convergent for linear approximators under ``a two-timescale framework, where the
main network is updated faster than the target network.'' \citet{Zhang:2023} provided sufficient conditions for
convergence of a SGD implementation of DQN, i.e. FVI using deep neural network  approximators $Q_\phi$ but using 
SGD iterations for $\phi_t$. Their analysis also 
allows for {\it experience replay,\/} i.e. the innovation of storing simulated outcomes in a replay buffer to help improve
stability and reduce the computational burden of SGD by using random sampling of simulated outcomes in ``minibatches'' 
from the replay buffer to reduce instabilities due to serially correlated training data and  smooth the learning process.

The DQN algorithm of \citet{Zhang:2023} involves a double loop of iterations, with $t=0,\ldots,T$ indexing the ``outer''
FVI iterations and $m=1,\ldots,M$ indexing the ``inner'' SGD iterations. Let $\phi_{t,m}$ be the parameter
value at iteration $(t,m)$ of this algorithm. The target network parameters are set to $\phi_{t,0}$ and are only
updated in the outer $t$ loop but are held fixed for all iterations $m$ in the inner SGD loop, whereas the inner
SGD iterations at FVI iteration $t$ are given by
\begin{equation}
     \phi_{t,m+1}=\phi_{t,m}+\alpha_t \left[\sum_{n \in \mathcal{D}_{t,m}} \left[r(s_n,a_n)+\beta \max_{a' \in A(s_{n+1})}
Q_{\phi_{t,0}}(s_{n+1},a')-Q_{\phi_{t,m}}(s_n,a_n)\right]\nabla_\phi Q_{\phi_{t,m}}(s_n,a_n)\right],
\label{eq:dqn_algorithm}
\end{equation}
where $\mathcal{D}_{t,m}$ is a {\it minibatch\/} of simulated trajectories drawn at random from 
the experience replay buffer $\mathcal{D}_t$ of trajectories $\{(s_n,a_n)|n=1,\ldots,N\}$ simulated for $N$ time steps
under a $\varepsilon_t$-greedy policy. The experience replay buffer $\mathcal{D}_t$ is refreshed at each step $t$ of the outer FVI
loop of the training process, but held fixed over the $M$ inner SGD iterations. Theorem 1 of \citet{Zhang:2023}
provides sufficient conditions (notably a realizability assumption that the true $Q$ lies in the
deep-neural-network class $Q_\phi$, their Assumption 1) for the DQN algorithm (\ref{eq:dqn_algorithm}) to generate estimates of $Q$ that
converge uniformly at rate $1/\sqrt{N}$, the size of the experience replay buffer: $\|Q_{\phi_{t,0}}-Q\|_\infty=O_p(1/\sqrt{N})$.\footnote{\footnotesize The sup-norm consistency of \citet{Zhang:2023} relies on realizability. \citet{vdlkallus:2026} relax this for fitted-Q evaluation: without Bellman completeness, reweighting each Bellman regression by the target-to-behavior stationary density ratio yields finite-sample convergence, in the $L_2$ norm of the target's stationary distribution, to the fixed point of the projected Bellman operator rather than sup-norm convergence to the true $Q$.}

As a result of these algorithm design features that mitigate the deadly triad, DQN is not only convergent 
but has been successfully applied to solve a number of important large scale problems that were previously intractable.
\citet{mnih:2015} used a DQN to train effective strategies
for classic Atari 2600 video games. Similar methods were subsequently used for
board games such as chess, Go and Shogu, see \citet{Silver:2018}. In these applications, the transition probability
$p$ for the states is unknown due to the unknown response of opposing players. However 
they obtained a model-free implementation of DQN  by simulating game paths using {\it self-play\/} where
opponent players make choices according to the same $\varepsilon_t$-greedy policy used by the player
who is being trained. \citet{Zhang:2023} cites additional empirical successes of DQN in robotics and autonomous
vehicles.

\section{Structural Estimation of DDCs\label{sec:ddc}}

The term ``structural'' refers to the attempt to infer the underlying preferences and beliefs of economic
agents rather than simply summarizing their behavior, the main goal of reduced-form econometrics.
In a DDC model the agent's decision rule $\pi(a|s)$ is referred to as a ``reduced-form model'' that captures behavior: structural econometrics 
refers to methods that attempt to invert $\pi$ to infer the underlying preferences and beliefs $\{\beta,r,p\}$ that generate observed behavior. 
Key references include \citet{rust:1987}, \citet{rust:1994} and 
the survey by \citet{AguirregabiriaMira:2010}.\footnote{\footnotesize See \citet{rust:2014} for a discussion 
of the origins of SE, dating back to work at the Cowles Foundation in the 1940s and the rationale for this literature including
the argument of \citet{lucas:1976} that structural models are superior to reduced-form for counterfactual prediction of policy changes.
The development of DP in the 1950s and 60s 
revolutionized economics by enabling models of sequential
decision making under uncertainty, dynamic strategic interactions, and dynamic equilibria 
in markets which are key features of nearly all economic problems.  DP led to new dynamic economic theories
based on {\it rational expectations\/} that
showed how beliefs about uncertain future events affect the current choices of rational forward looking agents.}

An obstacle to inference is the fact that the optimal policy $\delta^*$ in (\ref{eq:bellman_policy}) is 
a pure strategy (i.e. deterministic function of $s$).  This leads to a problem of {\it statistical degeneracy\/} because
different individuals in the same observed state $s$ will typically make different choices, something the model cannot explain. 
To explain this heterogeneity we  augment the
state variable as $x=(s,\varepsilon)$  where
$\varepsilon$ is a vector of idiosyncratic preference shocks observed by the agent but not by others.
Even though $\delta^*(s,\varepsilon)$ is still a pure strategy from the agent's standpoint, it produces choices that appear
random from the observer's standpoint. Under additional assumptions that include the assumption $A(s)$ is finite for all $s \in S$,
we can derive a {\it conditional choice probability\/} (CCP) $\pi(a|s)$ implied by the 
strategy $\delta^*$ that admits positive probabilities for all feasible choices $a \in A(s)$. Using this, we can form a likelihood
function to infer $\{\beta,r,p\}$ from data on agents' states and choices.

\citet{rust:1987} extended \citet{mcfadden:1973}'s work on static random utility models of discrete choice 
to DDC models, resulting in an analytic formula for the CCP $\pi(a|s)$  known as 
the {\it multinomial logit model\/} (``softmax'' function in the RL literature).
Suppose that for each $s$, $\varepsilon$ is a vector with the same length as $|A(s)|$ the number of actions, so for
$\varepsilon(a)$ is the component of $\varepsilon$ associated with action $a \in A(s)$. Assume that 
the reward for action $a$ in state $(s,\varepsilon)$ can be written as $r(s,a)+\varepsilon(a)$. 
We also assume that the transition density $p(s',\varepsilon'|s,\varepsilon,a)$ factors as $g(\varepsilon'|s')p(s'|s,a)$ 
for a conditional density $g(\varepsilon'|s')$, where $\varepsilon$ has a multivariate Gumbel or Type 1 extreme value (EV) distribution 
with common scale parameter $\sigma$ and expectation equal to $E\{\varepsilon(a)|s\}=0$ for $a \in A(s)$. 
Let $V(s,\varepsilon)$ be the optimal value of this MDP, the solution to the Bellman equation (\ref{eq:bellman}). 
Under these assumptions we can show $V(s,\varepsilon)$ has the representation
\begin{equation} V(s,\varepsilon)=\max_{a \in A(s)}[Q(s,a)+\epsilon(a)],
\label{eq:Vrep}
\end{equation}
where $Q=\Lambda_\sigma(Q)$ is the unique fixed point to the ``soft Bellman operator'' $\Lambda_\sigma$ given by
\begin{equation}
    \Lambda_\sigma(Q)(s,a) = r(s,a) + \beta \sum_{s'} \sigma \log\left(\sum_{a' \in A(s')} \exp\{Q(s',a')/\sigma\}\right)p(s'|s,a).
\label{eq:smoothQ}
\end{equation}
The derivation of (\ref{eq:Vrep}) and (\ref{eq:smoothQ}) use an important
property of the EV distribution for $\varepsilon$, namely that the expectation of $V(s,\varepsilon)$ with respect to $\varepsilon$
has an  analytical expression given by the so-called log-sum or ``soft max function'' $V(s)$ 
\begin{equation}
    V(s) \equiv   E\left\{V(\tilde s,\tilde \varepsilon)\left|\tilde s=s\right\}\right. 
    =  \int_\varepsilon \max_{a \in A(s)} \left[Q(s,a)+\varepsilon(a)\right]g(d\varepsilon|s) 
    =  \sigma \log\left(\sum_{a\in A(s)} \exp\{Q(s,a)/\sigma\}\right).
\label{eq:smoothV}
\end{equation}
Using equations (\ref{eq:smoothQ}) and (\ref{eq:smoothV}) we can derive a ``soft Bellman equation'' for $V$
\begin{equation}
    V(s)= \sigma \log\left(\sum_{a\in A(s)} \exp\left\{[r(s,a)+\beta \sum_{s'} V(s')p(s'|s,a)]/\sigma\right\}\right).
 \label{eq:smoothBellman}
\end{equation}
Equation (\ref{eq:smoothQ}) defines a ``soft $Q$'' as the fixed point of the smoothed Bellman operator
$Q_\sigma=\Lambda_\sigma(Q_\sigma)$, and (\ref{eq:smoothBellman}) defines $V$ as a fixed point of the smoothed 
Bellman operator 
$V_\sigma=\Gamma_\sigma(V_\sigma)$ and both are contractions. These operators are equivalent  to the corresponding 
``hard operators'' $\Lambda_0$ and $\Gamma_0$ in MDPs without the state variable
$\varepsilon$, and one can show that $Q_\sigma$ and $V_\sigma$ converge to the standard hard 
$Q$ and $V$ functions as $\sigma \downarrow 0$. 

The representation of $V(s,\varepsilon)$ in equation (\ref{eq:Vrep}) means that the decision rule $\pi(a|s,\varepsilon)$ 
is isomorphic to a static random utility model where $Q(s,a)+\varepsilon(a)$ is the reward from action $a$.
The CCP $\pi(a|s)$ is the conditional probability that the DM chooses action $a \in A(s)$
given the observed state $s$, and under the assumptions above, it has the multinomial logit form
\begin{eqnarray}
     \pi(a|s)  & =& \int I\left\{Q(s,a)+\varepsilon(a) \ge \max_{a' \in A(s)}[Q(s,a')+\varepsilon(a')]\right\}F(d\varepsilon|s)\nonumber \\
 &=&{ \exp\{Q(s,a)/\sigma\} \over \sum_{a' \in A(s)} \exp\{Q(s,a')/\sigma\}}.
\label{eq:ccp}
\end{eqnarray}
As $\sigma \downarrow 0$, $\pi(a|s)$ converges to a degenerate probability that equals 1 if 
$Q(s,a)$ is the largest $Q$ value  and 0 otherwise, i.e. to the optimal pure strategy $\pi$
 in equation (\ref{eq:bellman_policy}) for MDPs without the  $\varepsilon$ variable.
It follows that the choice-specific values $Q(s,a)$ in (\ref{eq:smoothQ}) and CCPs $\pi(a|s)$ in (\ref{eq:ccp})
are exactly the same as the ``soft Q'' values and optimal policy derived in the RL
literature under the assumption that agents care about entropy of the policy $\pi$, given in equation
(\ref{eq:vdefentropy}) of section~\ref{section:rl}.

\citet{AguirregabiriaMira:2002} showed how to implement a ``soft'' analog of {\it policy iteration,\/} see section~\ref{section:pi},
to solve the soft Bellman equation (\ref{eq:smoothBellman}). 
Equation (\ref{eq:ccp}) is the {\it policy improvement step,\/} the analog of (\ref{eq:bellman_policy}), that computes the  policy 
$\pi$ implied by $Q$. The  analog of the {\it policy valuation\/} step (\ref{eq:policy_valuation}) is 
the function $V_\pi$ that solves the linear system
\begin{equation}
   V_\pi(s) =\sum_{a \in A(s)} \pi(a|s)[r(s,a)+E_\pi\{\varepsilon(a)|s,a\}+\beta \sum_{s'} V_\pi(s')p(s'|s,a)],
\label{eq:smoothpolicyvaluation}
\end{equation}
where $E_\pi\{\varepsilon(a)|s,a\}$ is the conditional expectation of $\varepsilon(a)$ given that action $a$ is chosen
in state $s$  under decision rule $\pi$. The EV assumption implies that
$E_\pi\{\varepsilon(a)|s,a\}=-\sigma \log(\pi(a|s))$, and using equation (\ref{eq:ccp}), we see that right side of 
(\ref{eq:smoothpolicyvaluation}) equals the soft max function defining the soft Bellman operator in equation
(\ref{eq:smoothBellman}). Soft policy iteration solves the soft Bellman equation $V_\sigma=\Gamma_\sigma(V_\sigma)$ 
by cycling between policy valuation steps (\ref{eq:smoothpolicyvaluation})
and policy improvement steps (\ref{eq:ccp}), using the $Q_\pi$ implied by $V_\pi$ in equation (\ref{eq:smoothQ}). 
Hard/soft policy iteration is equivalent to using Newton's method to solve the hard/soft Bellman equations for $V$.

Equation (\ref{eq:smoothpolicyvaluation}) is  the Bellman equation for the entropy-adjusted MDP problem, (\ref{eq:vdefentropy}), where
the entropy term $H(\pi(s))$ equals  the expectation of the unobserved component of rewards $\varepsilon$ under an optimal pure strategy
$\delta^*(s,\varepsilon)$ for the MDP with augmented state variable $(s,\varepsilon)$, value function $V(s,\varepsilon)$ (\ref{eq:Vrep}), 
and implied CCP $\pi$ (\ref{eq:ccp}). It also implies that $\sigma\log(\pi(a|s))=Q(s,a)-V(s)\equiv \mathcal{A}(s,a)$, where
$\mathcal{A}(s,a)$ is known as the {\it advantage function\/} in the RL literature. We can use it to rewrite the fixed point
equation (\ref{eq:smoothQ}) for $Q$ as a linear system depending on $r(s,a)$ and $\log(\pi(a|s))$,
\begin{equation}
    Q(s,a) = r(s,a) + \beta \sum_{s'} \sum_{a' \in A(s')} \left[ -\sigma\log(\pi(a'|s'))+Q(s',a')\right]\pi(a'|s')p(s'|s,a).
\label{eq:linq}
\end{equation}
$Q$ decomposes as $Q(s,a)=Q_r(s,a)+Q_\varepsilon(s,a)$ where $Q_r(s,a)$ is the component 
of discounted value from observed rewards $r$ and $Q_\varepsilon(s,a)$ is the component from unobserved rewards $\varepsilon$.
Substituting expression (\ref{eq:smoothpolicyvaluation}) in place of (\ref{eq:smoothBellman}) in the equation for
$Q$ in (\ref{eq:smoothQ}), we can show that $Q_r$ and $Q_\varepsilon$ are solutions to the linear systems
\begin{eqnarray}
     Q_r(s,a)&=&r(s,a)+\beta \sum_{s'} \sum_{a' \in A(s')} Q_r(s',a')\pi(a'|s')p(s'|s,a) \label{eq:Qr} \\
     Q_\varepsilon(s,a)&=&\beta \sum_{s'} \sum_{a' \in A(s')}\left[-\sigma\log(\pi(a'|s'))+ Q_\varepsilon(s',a')\right]\pi(a'|s')p(s'|s,a).
\label{eq:Qeps}
\end{eqnarray}
The next section shows how this decomposition can reduce the computational burden of estimating DDCs.

\subsection{Estimation Methods for DDCs}
\label{section:ddc_methods}

The most common estimation method for DDCs is some form of  maximum likelihood.
Suppose we observe a random sample of {\it trajectories\/} $\mathcal{D}=\{\tau_i|i=1,\ldots,N\}$ 
where the trajectory $\tau_i$ for individual $i$ is a sequence of consecutive states and actions 
$\tau_i=\{(s_{it},a_{it})|t=0,\ldots,T_i\}$. Suppose  
that we smoothly parametrize $(r,p)$ using a vector unknown parameters $\theta$, which we denote by
$r_\theta$ and $p_\theta$.\footnote{\footnotesize
The discount factor  $\beta$ is often treated as known, though in 
some cases it is also estimated and thus part of $\theta$.}  
Using the implicit function theorem, we can show that $V_\theta$ and $Q_\theta$, the contraction fixed points given in equations
(\ref{eq:smoothV}) and (\ref{eq:smoothQ}), are continuously differentiable function of $\theta$.
Using these fixed points and the logit formula for $\pi(a|s)$ in equation (\ref{eq:ccp}) we can  write a full  likelihood
function $\mathcal{L}^f_{\mathcal{D}}(\theta)$ for the data as a function of the unknown parameters $\theta$ given by
\begin{equation}
    \mathcal{L}^f_{\mathcal{D}}(\theta) = \prod_{i=1}^N \prod_{t=1}^{T_i}  \pi_\theta(a_{it}|s_{it})p_\theta(s_{it}|s_{it-1},a_{it-1}).
\label{eq:lf}
\end{equation}
The smoothness of $V_\theta$ and $Q_\theta$ in $\theta$ implies that the CCP $\pi_\theta$ and the
likelihood $\mathcal{L}^f_{\mathcal{D}}(\theta)$ are both smooth functions of $\theta$ so the standard asymptotic
efficiency properties of parametric maximum likelihood estimation apply. 

The Nested Fixed Point algorithm (NFXP) \citep{rust:1988} computes the MLE of $\theta$ via a standard outer hill-climbing loop that maximizes $\mathcal{L}^f_{\mathcal{D}}(\theta)$, where each evaluation calculates
the contraction fixed point $Q_\theta=\Lambda_\theta(Q_\theta)$ that defines the choice-specific values $Q_\theta$ where $\Lambda_\theta$
is the operator defined on the right hand side of equation (\ref{eq:smoothQ}).
Notice that $\mathcal{L}^f_{\mathcal{D}}(\theta)$ depends on $\theta$ via $Q_\theta$. An alternative strategy \citep{sujudd:2012}
computes the MLE via the MPEC algorithm, maximizing $\mathcal{L}^f_{\mathcal{D}}(\theta,Q)$ as a function of $(\theta,Q)$
subject to the fixed point constraint $Q_\theta=\Lambda_\theta(Q_\theta)$.\footnote{\footnotesize \citet{sujudd:2012} claim that the
MPEC algorithm is substantially faster than NFXP, but this was due to using SA \citep{sujuddcomment:2016}
to compute the fixed point $Q_\theta=\Lambda_\theta(Q_\theta)$, which is slow when $\beta$ is close to 1. When Newton's method is used
to compute this fixed point, the cpu times for MPEC and NFXP are roughly comparable with NFXP being faster on test problems with
large sample sizes.}
NFXP's computational burden stems from computing the inner fixed point $Q_\theta=\Lambda_\theta(Q_\theta)$
for each trial value of $\theta$ in the outer hill-climbing algorithm.\footnote{\footnotesize \citet{oguz:2026} avoid the inner fixed-point loop via {\it unnested fixed point\/} (UFXP and OUFXP) estimators built on a dual representation of the Bellman equation, reporting DDC estimation speedups of up to three orders of magnitude while allowing neural-network utilities.} 
Newton's method (e.g. policy iteration) is typically used to solve for $Q_\theta$ or for $V_\theta$ from 
the soft Bellman equation (\ref{eq:smoothBellman}). Even though  Newton's method converges rapidly, it requires $O(|S|^3)$ operations per Newton/policy
iteration step, making NFXP burdensome for large state spaces $S$. Therefore alternative estimation methods have been proposed that 
reduce the total number of policy iteration steps required to estimate $\theta$.

The ``CCP estimator'' \citep{hotzmiller:1993}
for $r$ uses a non-parametric estimate of $\pi(a|s)$ to avoid repeated
solution of $Q$ over the search for $\theta$ that NFXP requires. Consider the case where $r_\theta$ is linear in parameters,
i.e. we can write $r_\theta(s,a)=r(s,a)*\theta \equiv \sum_{k=1}^K r_k(s,a)\theta_k$ for known functions
 $\vec{r}=(r_1,\ldots,r_K)$ called {\it features\/} in the IRL
literature. Then (\ref{eq:Qr}) implies that $Q_r(s,a)=R(s,a)^{\top}\theta$ where
$R$ is the solution to
\begin{equation}
   R(s,a)=\vec{r}(s,a) + \beta \sum_{s'} \sum_{a' \in A(s')} R(s',a')\pi(a'|s')p(s'|s,a) = \vec{r}(s,a)+\beta ER(s,a),
\label{eq:hequation}
\end{equation}
and $ER(s,a)$ is a shorthand for the conditional expectation of $R(s,a)$ in the middle of the expression in (\ref{eq:hequation}).
It follows that we only need to solve (\ref{eq:hequation}) once using a non-parametric estimate of $\hat\pi$ along with
$Q_\varepsilon$ in (\ref{eq:Qeps}) to serve as a ``correction term''  to obtain the following form for the CCP
\begin{equation}  
           \pi_\theta(a|s) = {\exp\{\hat R(s,a)^{\top}\theta+\hat Q_\varepsilon(s,a)\} \over \sum_{a' \in A(s)}
\exp\{\hat R(s,a')^{\top}\theta+\hat Q_\varepsilon(s,a')\}},
\label{eq:ccpestimator}
\end{equation}
where $\hat R(s,a)$ is the solution to (\ref{eq:hequation}) and $\hat Q_\varepsilon$ is the solution
to (\ref{eq:Qeps}) using non-parametric first stage
estimators $\hat\pi(a|s)$ and $\hat p(s'|s,a)$. The CCP estimator $\hat\theta$ maximizes the partial likelihood 
$\mathcal{L}^p_{\mathcal{D}}(\theta)$ given by 
\begin{equation}
    \mathcal{L}^p_{\mathcal{D}}(\theta) = \prod_{i=1}^N \prod_{t=1}^{T_i}  \pi_\theta(a_{it}|s_{it}),
\label{eq:ccplf}
\end{equation}
where $\pi$ is given in (\ref{eq:ccpestimator}).\footnote{\footnotesize See \citet{lrspe:2022} 
for  a modified version of the CCP estimator that is ``locally robust'' i.e. it corrects for sampling
error in the first stage estimates of the functions $\hat H(s,a)$ and $Q_\varepsilon(s,a)$.}
Thus, when $r$ is a  linear combination of known features, we only  need to compute $\hat R$ and $\hat Q_\varepsilon$ once
at the start of the estimation, requiring  $K+1$ solutions of systems with $|S|$ equations and unknowns,
where $K$ is the number of features. This can be further reduced to
a single solution if we assume that $r(s,a_s)$ is known for some
``normalizing alternative'' $a_s \in A(s)$ (see Section~\ref{section:identification}).

A cost of the CCP estimator is that it is less efficient than using NFXP to estimate the same partial likelihood criterion (\ref{eq:ccplf})
due to the noise in the first stage non-parametric estimates $\hat\pi$ that are used to construct
the $\hat Q_\varepsilon$ function in (\ref{eq:ccpestimator}).  \citet{AguirregabiriaMira:2002} (AM) addressed this problem
by proposing an iterative estimator called {\it nested pseudo likelihood\/} (NPL)  that uses an initial non-parametric estimate $\hat\pi$ 
to compute the value function $V_{\hat\pi}$ implied by this initial estimate by solving for it in the policy valuation step
in equation (\ref{eq:smoothpolicyvaluation}), which we denote as $\hat V_{\theta,\hat\pi}$ since it depends
both on $\hat\pi$ and the parameters $\theta$ entering
the reward function. NPL  estimates  $\hat\theta$ using a pseudo-likelihood similar to (\ref{eq:ccplf})
except that the CCP $\pi_\theta(a|s)$ is evaluated using (\ref{eq:ccp}) using estimated $Q$ functions given by
\begin{equation}
                  \hat Q_\theta(s,a)=r_\theta(s,a)+\beta \sum_{s'}\hat V_{\theta,\hat\pi}(s'|s,a)p(s'|s,a),
\label{eq:qhat}
\end{equation}
Thus, the likelihood $L_\pi(\theta)$ depends implicitly on the first stage non-parametric CCP estimate $\hat\pi$, which can be regarded
as a ``nuisance parameter''. In principle, estimation noise in $\hat\pi$ will contaminate and affect the asymptotic 
covariance matrix for the parameters of interest $\theta$. However 
at the fixed point, we have $\partial V(s)/\partial \pi=0$ and AM showed that this  {\it zero Jacobian property\/}  
implies that  the information matrix for the parameters of interest $\theta$ 
is ``Neyman orthogonal'' with respect to estimation error in the first stage nuisance parameter $\hat\pi$, 
so the NPL estimator of $\theta$ is unaffected by noise in the first stage $\hat\pi$, at least asymptotically.
In smaller samples the orthogonality may not hold exactly, but 
NPL can be iterated to result in improved finite sample properties.
An iterative or $k$-step version of NPL
uses the initial estimate  $\hat\theta$ to update the estimate of the CCPs using equation (\ref{eq:ccp}), then using the updated policy
$\hat\pi=\pi_{\hat\theta}$ and a soft policy valuation step (\ref{eq:linq}) to obtain
updated $V$ and $Q$ functions (\ref{eq:qhat}),  which are in turn
used in the likelihood (\ref{eq:ccplf}) to produce an updated estimate $\hat\theta'$. Iterative NPL amounts to ``swapping''
the policy iteration and maximization steps of the NFXP estimator (NFXP  maximizes the likelihood over $\theta$ in the ``outside'' loop
and does policy iteration in the ``inside'' loop). Swapping the order of maximization
and policy iteration reduces the total number of policy iteration steps required to 
estimate $\theta$. AM showed that if the iterated version of  NPL converges, it produces the
same  estimate $\hat\theta$ that  NFXP produces from maximization of the partial likelihood  (\ref{eq:ccplf}).
Thus the iterated NPL estimator is as efficient but faster than NFXP and more efficient but slower than the CCP
estimator.\footnote{\footnotesize
See also {\it efficient pseudo likelihood\/} (EPL) \citep{dearingblevins:2025}, a similar iterative
procedure that is also an efficient sequential estimator that 
can estimate parameters of dynamic games.}  

The extreme value (EV) assumption, while restrictive, dramatically reduces computational burden 
via the analytic expressions it implies for the softmax CCP (\ref{eq:ccp}) and expected value function (\ref{eq:smoothV}).
Other natural choices for the unobserved $\varepsilon$ such as multivariate normal do not result in analytical expressions  
and thus considerable computation to approximate these quantities. 
The conditional independence (CI) of the $\{\varepsilon_t\}$ shocks is another restrictive assumption driven by
computational considerations. Though there are computationally
efficient recursive likelihood integration methods, e.g. \citet{reich:2018}, that can allow
for serial correlation in $\varepsilon_t$ shocks, these methods are still burdensome for estimation.
\citet{ijc:2009} and \citet{norets:2009} introduced
Bayesian inference in DDC models that use Markov Chain Monte Carlo (MCMC) methods to
simulate from rather than directly compute the posterior by numerical integration. 
MCMC facilitates {\it data augmentation\/} where latent variables 
can be simulated symmetrically with observed ones rather than being marginalized from the likelihood, 
allowing us to relax the EV and CI assumptions. They also show that RL can be used to update the Q functions
of the underlying MDP problem in parallel inside the overall do-loop for the MCMC simulations
where the structural parameters are drawn from the likelihood using Gibbs' or Metropolis-Hastings' samplers.

There are also non-Bayesian, non-likelihood based  simulation estimators such as the {\it method of simulated moments\/}
(MSM) \citet{msm:1989}, {\it indirect inference\/} \citet{ii:1993},  the {\it moment inequality\/} estimator of
\citet{bbl:2007}, and the {\it adversarial estimator\/} of \citet{kaji:2023} 
that use stochastic simulations to  estimate problems where it is difficult to explicitly specify or compute a likelihood function. 
All the estimators discussed so far are designed for problems where the state space is discrete and small enough to be enumerable. When the state space is large or contains continuous variables, these methods become computationally infeasible, motivating the function approximation techniques we turn to next.

\subsection{Function Approximation Methods for Large State Spaces}

When the state space $S$ is large (i.e. the DDC has many continuous state variables that must be discretized)
the approaches discussed above that use PI and repeated solution of
large linear systems with $|S|$ equations and unknowns are computationally infeasible. 
Following section~\ref{section:curse} we can approximate 
the fixed point $Q=\Lambda_\sigma(Q)$ of the soft Bellman operator in equation (\ref{eq:smoothQ}) by $Q_\phi(s)$
with a flexible parametric family such as a neural network with weights $\phi$, and  approximate
the solutions to 
the fixed point problems (\ref{eq:smoothQ}) and (\ref{eq:smoothV}) more tractably as nonlinear least squares problems.  
Let $\mbox{BE}(\phi,\theta)$ be the mean squared Bellman error defined in equation (\ref{eq:dqnideal}).
It depends both on the parameters (weights) $\phi$ of the neural network approximation $Q_\phi$
as well as the structural parameters $\theta$ via the soft Bellman operator $\Lambda_{\sigma,\theta}(Q_\phi)$ 
defined in (\ref{eq:smoothQ})  via the dependence of the structural functions $\{\beta,r,p\}$ on the parameters
$\theta$.  A nested least squares estimator similar to NFXP can estimate $\theta$ 
by maximizing the likelihood $\mathcal{L}^f_{\mathcal{D}}(\theta)$ (\ref{eq:lf}) while minimizing 
$\mbox{BE}(\phi,\theta)$ over $\phi$ for
each value of $\theta$. A penalized likelihood estimator called SEES \citep{LuoSang:2024}, for {\it sieve-based efficient estimator for
structural models\/}, chooses $\hat\theta$ and $\hat\phi$ to maximize 
\begin{equation}
(\hat\theta,\hat\phi)=\argmax_{\theta,\phi} \log(\mathcal{L}^f_{\mathcal{D}}(\theta))-\lambda \mbox{BE}(\phi,\theta),
\label{eq:penalizedllf}
\end{equation}
where $\lambda \ge 0$ is a penalization parameter. Consistency of SEES requires the use of a {\it sieve,\/} i.e.
the dimension of $\phi$ must grow to infinity at certain rate with the sample
size so that the approximation error in approximating $Q$ by $Q_\phi$ converges to zero.
A related estimator, the {\it neural network efficient estimator\/} (NNES) of \citet{Nguyen:2025}, maximizes a penalized criterion
similar to SEES in a modified version of the sequential NPL
estimator of \citet{AguirregabiriaMira:2002}.

A challenge facing sieve-based estimators is that $\phi$ are high dimensional  ``nuisance parameters'' 
(their only purpose is to approximate $Q$) whereas $\theta$ are the ``parameters of interest'' 
that determine  rewards and beliefs, $\{\beta,r,p\}$. Since $\hat\phi$  minimizes
the sum of squared residuals $\mbox{BE}(\phi,\theta)$  for each $\theta$, it follows that $\hat\phi$ will be an implicit
function of $\theta$. The asymptotic covariance matrix for $\hat\theta$  requires $\nabla_\theta Q_{\hat\phi(\theta)}$, the gradient
of $Q_{\hat\phi(\theta)}$ with respect to $\theta$. $\hat\phi(\theta)$ is a well defined smooth function
of $\theta$ by the Implicit Function Theorem, and differentiating 
the first order condition for $\hat\phi(\theta)$, 
$\nabla_\phi Q_\phi=0$ at $\phi=\hat\phi(\theta)$, to compute $\nabla_\theta \hat\phi(\theta)$  requires 
calculating the Hessian matrix $\nabla^2_{\phi,\phi} Q_\phi$  which is typically infeasible to compute.
Thus computing $\nabla_\theta Q_{\hat\phi(\theta)}=\nabla_\phi Q_{\hat\phi(\theta)}\nabla_\theta\hat\phi(\theta)$
via the chain rule is generally out of the question. Computing $\nabla_\theta Q$ via numerical differentiation of
$Q_{\hat\phi(\theta)}$ with respect to $\theta$ is also time-consuming and potentially inaccurate
because  multiple local optima in $\phi$ result in fragile
solutions $\hat\phi(\theta)$ where small changes in $\theta$ produce discontinuous jumps in $\hat\phi(\theta)$ and $Q_{\hat\phi(\theta)}$. 
\citet{Nguyen:2025} shows that a fast and accurate way to compute $\nabla_\theta Q_{\hat\phi(\theta)}$ is to use the
method of {\it equilibrium propagation\/} introduced by \citet{ep:2022}.

\subsection{The Identification Problem}
\label{section:identification}

Section~\ref{sec:mdp} showed how $\pi$ is calculated from the structure $\{\beta,r,p\}$, which
we summarize by the mapping $\pi=f(\beta,r,p)$. The Identification Problem asks if this mapping is invertible:
given $\pi$ is there a unique underlying structure $\{\beta,r,p\}$ that implies it? Unfortunately, without
further restrictions the answer is negative, as shown by \citet{hotzmiller:1993}, 
\citet{rust:1994}, \citet{nhr:1999}, and \citet{magnacthesmar:2002}. 
Consider the first step of the inversion process, uncovering  $Q$ from $\pi$:
\begin{equation}
               \log\left(\pi(a|s)/\pi(a'|s)\right) = Q(s,a)-Q(s,a'), \quad s \in S, \; a,a' \in A.
\label{eq:ccpinverse}
\end{equation}
Thus, knowledge of $\pi$ is insufficient to recover all the $Q$ values: only the {\it differences\/} in
$Q$ are identified. Indeed, the structure $\{\beta,r,p\}$ contains $1+|S|(|A|-1)+|S|^2|A|$  
parameters, more than the $|S|(|A|-1)$ parameters of $\pi$, so we will generally have ``more equations than unknowns'' 
so the structure is only partially identified. Unique (or ``point'') identification requires additional restrictions on $\{\beta,r,p\}$. 
A common assumption is that $r$ and $p$ have known parametric functional forms that depend on a relatively small number of parameters $\theta$. In some situations the reward function can be treated as
known, and this facilitates the identification of subjective beliefs $p(s'|s,a)$.\footnote{\footnotesize 
For example \citet{tennis:2025} are able to identify the subjective beliefs of professional tennis servers about how their serve
strategies affects their probability of winning a tennis game. These beliefs take the form of a transition $p(s'|s,a)$ where $s$ denotes
the point state of the game and $a$ denotes serve direction. Identification of beliefs is possible  because it is reasonable to assume
$r$ is known: i.e. the server earns a reward of 1 if they win the game and 0 otherwise, and due to reasonable parametric restrictions
on $p$.} Rational expectations is a strong identifying assumption, 
i.e. agents' subjective beliefs $p(s'|s,a)$ correspond to the actual probability governing
these transitions. With sufficient data, $p$ can be estimated non-parametrically and treated as known from the standpoint of 
identification.  The discount factor $\beta$ is often assumed to be known {\it a priori,\/} though \citet{abbringdaljord:2020} showed that $\beta$ is identified
under rational expectations with certain restrictions on rewards.\footnote{\footnotesize \citet{yao:2024} show that observing experts with different planning horizons identifies both rewards and discount factors.}

When  $\{\beta,p\}$ are known, identification of $r$ depends on the assumption that either it has a known parametric
functional form $r_\theta$ where the dimension of $\theta$ 
is less than $|S|(|A|-1)$, or that for each $s \in S$ there is
a normalizing or ``anchor action'' $a_s$ such that  $r(s,a_s)$ is known. The latter assumption amounts to $|S|$ restrictions
that imply there are only $|S|(|A|-1)$ unknown rewards to be recovered, the same number of free values of $\pi$. 
\citet{hotzmiller:1993} and \citet{kang:2025} show that in this ``exactly identified'' case,
we can calculate $Q(s,a_s)$ from knowledge of $r(s,a_s)$ and $\pi(a_s|s)$ using equation (\ref{eq:linq}) 
and then (\ref{eq:ccpinverse}) allows us to recover all remaining $Q$ values. 
Using the $Q$ values we can back out the remaining unknown rewards $r$ from the soft Bellman equation (\ref{eq:smoothQ}).\footnote{\footnotesize 
\citet{eerl:2026} provide an alternative condition that implies exact identification of $r$, namely that there exists
some known policy $\pi(a|s)$ with zero expected reward, $0=\sum_{a \in A(s)} r(s,a)\pi(a|s)$, for all $s \in S$. }

The assumption that $r(s,a_s)$ is known for some normalizing action $a_s$ in each state $s \in S$ is far from innocuous. If wrong, estimated rewards will not equal true rewards even though the model perfectly fits $\pi$, and counterfactual predictions may differ from actual outcomes. 
For example we can always estimate rewards with the trivial estimator $\hat r(s,a)=\log(\pi(a|s))$ that perfectly fits
observed behavior and is optimal, i.e. the implied $Q$ function satisfies the fixed point $Q=\Lambda_\sigma(Q)$
in (\ref{eq:smoothQ}). However easy to see this estimator differs from the true reward: $\hat r(s,a)=r(s,a)+\beta \int V(s')p(s'|s,a)-V(s)$.
Indeed, in the absence of any prior restrictions on rewards, we can only identify a set of rewards 
$\mathcal{R}(r)=\{r_h\}$ that are observationally equivalent to $r$ given by $r_h(s,a)=r(s,a)+\beta \sum_{s'} h(s')p(s'|s,a)-h(s)$ for any function $h: S \to R$. 
\citet{nhr:1999} refers to  $\mathcal{R}(r)$ as ``potential-based reward shaping'' 
and shows that {\it policy invariance\/} holds: i.e.  $\pi_{r_h}=f(\beta,r_h,p)=\pi_r=f(\beta,r,p)$ for all
$r_h \in \mathcal{R}(r)$.  However for alternative transition probabilities $\lambda(s'|s,a)\ne p(s'|s,a)$ policy invariance
does not hold: $\pi_{r_h}=f(\beta,r_h,\lambda) \ne \pi_{r}=f(\beta,r,\lambda)$, i.e. counterfactual behavior 
implied by observationally equivalent rewards $r_h$ and $r$ differ in the new environment $\lambda$.

Identification can potentially be established if we have the luxury of being
able to conduct experiments with subjects to see how their behavior changes in different environments.
For example, \citet{cao_irl:2021} prove rewards are identifiable up to a constant if the agent is observed in two environments with sufficiently different transition laws.
The DDC and IRL literatures use different assumptions to overcome  the identification problem,
reflecting different goals of the respective fields. \citet{eerl:2026} and \citet{souza:2021} 
show that some counterfactuals are identified even if $\{\beta,r,p\}$ 
is partially identified, providing a  ``call for caution while leaving room for optimism: 
although counterfactual behavior and welfare can be sensitive to identifying restrictions imposed on the model, 
there exists important classes of counterfactuals that are robust to such restrictions.'' (p. 385).\footnote{\footnotesize \citet{rolland:2022} prove that linearly independent transition matrices across environments guarantee transfer even without full reward identification.}

\section{Inverse Reinforcement Learning\label{sec:irl}}

The IRL literature \citep[p.~663]{ng:2000} has two motivations for inferring rewards: 1) ``the potential use of reinforcement learning and
related methods as potential models for human and animal learning'' and 2) to provide a reward function
that RL can use to construct an ``intelligent agent that can behave successfully in a particular domain.'' Though
they note that imitation and apprenticeship learning (IL and AL) can be used to learn a policy $\pi$, ``the reward
function often provides a much more parsimonious description of behavior'' \citep[pp.~663--664]{ng:2000} so that IRL can constitute an effective
form of AL. IRL has become an important method in many areas of robotics and \citet[p.~5446]{av:2025} note that IRL has ``found great success in predicting 
trajectories through inferring cost or reward functions from expert demonstrations, 
and then using these functions to guide the behavior of self-driving vehicles in unseen driving environments.''

The IRL problem \citep{ng:2000} is to infer the reward function $r_\theta(s,a)$ from expert demonstrations. Given observed trajectories $\mathcal{D}=\{\tau_i\}_{i=1}^N$ where $\tau_i=\{(s_{it},a_{it})\}_{t=0}^{T_i}$, or an expert policy $\pi_E$ (the demonstrator's behavioral policy), we seek the parameters $\theta$ that rationalize observed behavior. The problem is ill-posed: many $\theta$ can yield the same observed behavior. For linear rewards, \citet{abbeel:2004} showed that matching the expert's expected discounted feature counts (the discounted sum of features accumulated along trajectories under a policy) suffices for policy imitation but does not uniquely identify $\theta$. Different IRL algorithms represent different principles for selecting a unique $\theta$ from this equivalence class. We selectively review methods most relevant to DDC and economic applications, referring readers to \citet{azar:2020}, \citet{arora:2021}, \citet{gleave:2022}, and \citet{adams:2022} for comprehensive surveys.

Originally, \citet{ng:2000} proposed selecting a unique $\theta$ via maximum margin: require the expert's cumulative reward to exceed any alternative policy's by a margin. \citet{ratliff:2006} refined this by making the margin grow with the alternative's deviation from expert behavior, penalizing vastly different policies more heavily.\footnote{\footnotesize This framework resembles moment inequality estimators in dynamic games \citep{bbl:2007}.} However, as \citet{ziebart:2010} argued, margin methods treat the expert demonstration as a clear ``winner'' that must beat every alternative by a fixed cushion under one reward function. Real demonstrations are noisy, sometimes suboptimal, and features inevitably miss factors the expert cared about. These limitations motivated the probabilistic approaches that follow.

\subsection{Maximum Entropy and Maximum Likelihood Methods}

The first use of likelihood-based approach in IRL with a softmax formulation for $\pi$ equal to the logit CCP (\ref{eq:ccp}) in the
DDC literature was Bayesian IRL (BIRL) \citep{birl:2007}. Rather than seeking a single reward function, BIRL treats IRL as Bayesian inference: it specifies a prior $\Pi(r)$ over reward functions, defines a likelihood of observed behavior given $r$, and uses MCMC to sample from the posterior $\Pi(r|\mathcal{D})$. The likelihood models expert actions using the softmax form (\ref{eq:ccp}), but with the ``hard $Q$'' associated
with the Bellman equation (\ref{eq:q1}) with unobserved states/shocks $\varepsilon$
rather than the internally consistent ``soft $Q$'' equation (\ref{eq:smoothQ}) that accounts for these shocks. 
Thus BIRL treats stochastic choice as optimization error rather than entropy-regularized optimality. \citet{birl:2007} showed that the maximum margin algorithm of \citet{ng:2000} corresponds to the maximum a posteriori (MAP) estimator under a Laplacian prior, revealing that margin maximization implicitly assumes sparse reward structures.\footnote{\footnotesize BIRL's local normalization produces {\it label bias\/}: paths with fewer branches receive higher probability even when total rewards are equal. MCE IRL eliminates this via soft Bellman backups that couple each state's value to all future branching. MCE also offers convex optimization (versus MCMC), a unique solution, robustness to suboptimal demonstrations, and natural extension to deep RL; see \citet{ni:2022} for a unified treatment.}

\citet{ziebart:2010} introduced {\it maximum causal entropy\/} (MCE) IRL, deriving the optimal policy from entropy maximization. Following \citet{jaynes:1957}, when multiple policies are consistent with observed behavior, maximum entropy selects the one least committed to structure beyond what the data require. Let $\mu_{\mathcal{D}} = \frac{1}{N}\sum_{i=1}^{N}\sum_{t=0}^{T_i} \beta^t \vec{r}(s_{it},a_{it})$ denote the expert's discounted feature counts. MCE IRL solves
\begin{equation}
    \max_{\pi} H_c(\pi) \quad \text{subject to} \quad E_\pi\left\{\sum_{t=0}^{T} \beta^t \vec{r}(s_t,a_t)\right\} = \mu_{\mathcal{D}},
\label{eq:mce_primal}
\end{equation}
where $H_c(\pi) = E_\pi\{-\sum_{t=0}^{T} \beta^t \log \pi(a_t|s_t)\}$ is causal entropy.\footnote{\footnotesize \citet{ziebart:2008} originally maximized Shannon entropy $H(\tau) = -\sum_\tau P(\tau)\log P(\tau)$ over trajectories $\tau$, which suffices for deterministic MDPs. However, Shannon entropy includes randomness from state transitions, creating a risk-seeking bias in stochastic MDPs. Causal entropy isolates policy randomness; the two coincide when transitions are deterministic.} Introducing Lagrange multipliers $\theta \in \mathbb{R}^K$ and solving the KKT conditions yields the softmax policy (\ref{eq:ccp}), where $Q_\theta$ and $V_\theta$ satisfy the soft Bellman equations (\ref{eq:smoothQ})--(\ref{eq:smoothV}) with linear reward $r_\theta(s,a) = \vec{r}(s,a)^{\top}\theta$.\footnote{\footnotesize \citet{pitombeira:2024} independently derive this softmax policy from RUM with Gumbel shocks, rediscovering the DDC approach.} This establishes mathematical equivalence between MCE IRL and DDC under extreme value (EV) shocks; \citet{ziebart:2010}'s maximum causal entropy IRL is nearly isomorphic to DDC models under the AS/CI/EV assumptions discussed in section 3.

Equivalently, assuming experts choose according to (\ref{eq:ccp}), maximum likelihood estimation yields the log partial likelihood (\ref{eq:ccplf}). The gradient takes the form $\nabla_\theta \log \mathcal{L}^p_{\mathcal{D}}(\theta) = \mu_{\mathcal{D}} - E_{\pi_\theta}\{\sum_{t} \beta^t \vec{r}(s_t,a_t)\}$, which is the difference between expert and model feature expectations. At the optimum these expectations match, precisely the constraint in (\ref{eq:mce_primal}), confirming equivalence of the entropy and likelihood perspectives \citep{ziebart:2010}.\footnote{\footnotesize This duality (that maximizing entropy subject to feature-matching constraints yields the same solution as maximum likelihood in exponential families) is a general result; see \citet{berger:1996} for a comprehensive treatment in the context of natural language processing. \citet{ziebart:2008} applied MCE IRL to taxi driver route choices in Pittsburgh, correctly predicting nearly 80\% of paths in a holdout sample.} \citet{mlirl:2011} made this MLE interpretation explicit by directly maximizing trajectory likelihood.

The Max-Ent IRL algorithm \citep{ziebart:2010} iterates until convergence. First, a backward pass solves the soft Bellman equation (\ref{eq:smoothQ})--(\ref{eq:smoothV}) for $Q_\theta$ and $V_\theta$, from which the policy $\pi_\theta(a|s)$ follows from (\ref{eq:ccp}). Next, a forward pass computes expected state-action visitation $E_{\pi_\theta}\{\sum_t \beta^t \vec{r}(s_t,a_t)\}$ via dynamic programming or Monte Carlo simulation. Finally, the gradient update $\theta \leftarrow \theta + \alpha(\mu_{\mathcal{D}} - E_{\pi_\theta}\{\cdot\})$ adjusts reward parameters. This requires solving the soft Bellman equation and computing state visitation frequencies at each iteration, making the method model-based. Classic MaxEnt IRL further requires a discrete, enumerable state space and knowledge of $p(s'|s,a)$, limiting applicability to tabular settings.\footnote{\footnotesize \citet{kim:2021} prove strong identifiability in tabular MaxEnt requires the MDP's domain graph to be coverable and aperiodic.}

\citet{guidedcostlearning:2016} (Guided Cost Learning, GCL) scaled MaxEnt IRL to high-dimensional continuous domains by addressing these limitations. Three key innovations enable this: (1) neural networks enable both feature learning and nonlinear cost representation, replacing hand-crafted features with more flexible reward functions $r_\theta(s,a)$; (2) importance sampling corrects the distribution mismatch when policy samples come from a suboptimal generator rather than the current optimal policy; and (3) lazy policy optimization runs the inner RL loop for only a few iterations per reward update, avoiding the prohibitive cost of full convergence. GCL approximates the MaxEnt gradient $\mu_{\mathcal{D}} - E_{\pi_\theta}\{\vec{r}\}$ using these sampled, reweighted trajectories, enabling end-to-end training in continuous control tasks.

\subsection{Adversarial Methods}

Adversarial methods frame IRL as a different type of bi-level optimization: an outer loop updates reward parameters while an inner loop finds the optimal policy, implemented via a discriminator-generator game. While Max-Ent IRL requires solving Bellman equations (and hence knowing $p(s'|s,a)$), adversarial methods offer a model-free alternative using the \emph{discounted occupancy measure} $d_\pi(s,a) = E_\pi\{\sum_{t=0}^\infty \beta^t \mathbb{I}(s_t=s, a_t=a)\}$, which reformulates cumulative reward as an inner product $\langle d_\pi, r \rangle$. \citet{ho:2016} show that relaxing feature-matching to occupancy-measure divergence yields Generative Adversarial Imitation Learning (GAIL), resulting in the saddle point problem
\begin{equation}
    \min_\pi \max_{D: S \times A \to (0,1)} E_{d_E}\{\log D(s,a)\} + E_\pi\{\log(1-D(s,a))\} - \lambda H_c(\pi),
\label{eq:gail_saddle}
\end{equation}
where $d_E$ is the expert's occupancy measure and $\lambda$ weights causal entropy $H_c(\pi)$. The inner maximization trains a discriminator $D$ to distinguish expert from policy samples; the outer minimization updates $\pi$ via RL to fool $D$. At equilibrium, the objective equals the Jensen-Shannon divergence between occupancy measures. GAIL is model-free: only samples from $\pi$ and the expert are needed. However, the optimal discriminator approximates the density ratio $D^*(s,a) = d_E(s,a)/[d_E(s,a) + d_\pi(s,a)]$, and because occupancy measures depend on transition dynamics $p(s'|s,a)$, the induced reward $\log D(s,a)$ bakes in training dynamics and causes policy failure when transferring to new environments.\footnote{\footnotesize \citet{fu:2018} demonstrate this dramatically: a GAIL-trained quadrupedal ant fails completely when two legs are disabled at test time, while AIRL's portable reward allows the agent to develop a new crawling gait.} This failure of transferability is a major drawback for counterfactual analysis. To recover a truly \emph{portable} reward, \citet{fu:2018} introduced Adversarial IRL (AIRL) with a structured discriminator whose logits decompose as
\begin{equation}
    f_\phi(s,a,s') = g_\phi(s,a) + \beta h_\phi(s') - h_\phi(s).
\label{eq:airl_decomp}
\end{equation}
Here $g_\phi: S \times A \to \mathbb{R}$ is the reward approximator and $h_\phi: S \to \mathbb{R}$ is a learned shaping potential, both parameterized by neural network weights $\phi$. This structure implements potential-based reward shaping from Section~\ref{section:identification}: $g_\phi$ captures the intrinsic reward while $h_\phi$ absorbs dynamics-dependent value contributions.\footnote{\footnotesize AIRL's per-transition decomposition $f_\phi(s,a,s')$ matches the original definition in \citet{nhr:1999} (Theorem~1): $F(s,a,s')=\gamma\Phi(s')-\Phi(s)$. The expected form $r_h(s,a)$ in Section~\ref{section:identification} integrates this over $p(s'|s,a)$.} The discriminator takes the form
\begin{equation}
    D_\phi(s,a,s') = \frac{\exp(f_\phi(s,a,s'))}{\exp(f_\phi(s,a,s')) + \pi(a|s)}.
\label{eq:airl_discriminator}
\end{equation}
After training, the portable reward is extracted as $r(s,a) = g_\phi(s,a)$.

The theoretical guarantee requires restrictions. Under deterministic dynamics, a reward function $r(s)$ that does not depend on action $a$, 
and $g_\phi$ restricted to depend only on $s$, \citet{fu:2018} prove that at the optimum $g_\phi(s) = r(s) + \text{const}$ and $h_\phi(s) = V(s) + \text{const}$. Thus $g_\phi$ recovers the intrinsic reward up to a constant, enabling transfer to new dynamics. Without the state-only restriction, or under stochastic dynamics, the decomposition is approximate. 

\citet{kaji:2023} provide theoretical foundations showing adversarial estimation is a principled extremum estimator. With a logistic discriminator distinguishing real from simulated data, the adversarial objective asymptotically reduces to optimally-weighted MSM, automatically selecting informative moments. Under correct specification the estimator achieves parametric efficiency; under misspecification it attains the parametric rate. When applied to IRL, the generator corresponds to $\pi_\theta$ producing occupancy measures while the discriminator plays the same role as in AIRL. This justifies why adversarial IRL scales to high-dimensional settings: the discriminator is an optimal moment selector that adapts to the data distribution, not merely a learned cost function. 

Recent work addresses AIRL's identification limitations in economic settings. The disentanglement theorem of \citet{fu:2018} requires deterministic dynamics and state-only rewards $g_\phi(s)$, assumptions rarely satisfied in economic applications where utilities are inherently action-dependent.\footnote{\footnotesize When rewards depend on actions, identification requires normalizing assumptions analogous to the anchor action in DDC (Section~\ref{section:identification}); see \citet{lee:2026} below.} When rewards depend on actions, the decomposition (\ref{eq:airl_decomp}) is not unique: variation in $g_\phi$ can be offset by changes in $h_\phi$, creating an entanglement between immediate payoffs and forward-looking anticipation. 
\citet[p.~1]{lee:2026} showed how to relax these restrictions in an empirical application to serialized content consumption (i.e. where consumers read fiction or
watch television series in successive episodes) which is inherently forward-looking: ``users weigh the immediate cost of access against the expected value of continuing further along the
sequence, with content shaping how those trade-offs evolve over time.''
They show that economic structure provides a credible identifying normalization to distinguish current and future rewards:
 consumers have an exit option with known flow payoff $g_\phi(s, \text{exit}) = 0$ leading to an absorbing state with $h_\phi(s_{\text{abs}}) = 0$. These anchors pin down the shaping potential, then $g_\phi(s, \text{read})$ is identified by substitution. This mirrors the normalizing action assumption in DDC identification \citep{rust:1994,magnacthesmar:2002}. Methodologically, \citet{lee:2026} demonstrate AIRL can be used with neural networks
to estimate reward functions over unstructured chapter text as high-dimensional state variables while allowing for unobserved
heterogeneity in consumers. Using the estimated reward functions  they evaluate counterfactual pricing policies such as
wait for free access and demonstrate that ``For high–willingness-to-pay users, free access facilitates progression
through weaker episodes and can increase downstream purchases'' and ``concentrating paywalls at
high-engagement episodes and leaving weaker episodes free preserves progression and raises monetization'' \citep[p.~33]{lee:2026}.

\subsection{Model-Free IRL}

The AIRL estimator of serialized fiction consumption is an example of model-free IRL. The economic model presumes individuals have 
a belief $p(s'|s,a)$ about how content of the next episode $s'$ depends on the current episode $s$, which is key to their binary decision $a$
of whether to consume the next episode, yet it is not clear how to specify or estimate this belief empirically.
As \citet[p.~4327]{adams:2022} notes, ``in many scenarios, the transition function may be unknown and difficult or impossible to estimate.''\footnote{\footnotesize In RL, ``model-free'' means using raw transitions $(s,a,s')$ without explicitly estimating $p(s'|s,a)$, whereas ``model-based'' means constructing or estimating $p$ for planning \citep{BartoSutton:2020}. In SE/DDC, methods using historical data are often called ``model-based'' since $p$ can be estimated from observed transitions. We follow RL terminology throughout.} Model-free IRL avoids this by using observed transitions $(s,a,s')$ directly, reminiscent of experience replay in deep Q-learning \citep{mnih:2015}. \citet{jain:2019} is an early example of model-free IRL, using a  MLE framework with a Max-Ent 
log-likelihood objective but replacing the model-based Bellman update with a differentiable variant of Q-learning, where
gradients are estimated using only sampled transitions. This section discusses two other MLE-based model-free approaches to IRL that
use function approximation to handle large state spaces.

\citet{AE:2022} (AE) proposed a model-free estimator that combines a core method of RL, the method of {\it temporal differences\/} (TD),
\citet{BartoSutton:2020}, and the two step \citet{hotzmiller:1993} CCP estimator in
section~\ref{section:ddc_methods} that avoids the need to estimate or make parametric assumptions about $p(s'|s,a)$. Similar to the CCP estimator, they
assume the reward is a linear combination of known features, i.e. $r_\theta(s,a)=\vec{r}(s,a)^{\top}\theta$,
and  estimate the structural parameters $\theta$ by maximizing the partial likelihood (\ref{eq:ccplf}). 
However rather than numerically computing functions $R$ and $Q_\varepsilon$ that
enter as covariates in the CCP $\pi$ in equation (\ref{eq:ccpestimator}), they use observed state transitions
to estimate  $R$ and $Q_\varepsilon$ under the assumption they can be approximated as linear
combinations of a set of $K$ known basis functions $\vec{\rho}=(\rho_1,\ldots,\rho_K)$ that depend on $(s,a)$ and a 
corresponding  vector of parameters $\phi$. Let $R_\phi(s,a) = \vec{\rho}(s,a)^{\top}\phi$
be the approximation of $R$, and assume a similar approximation is used to compute $Q_\varepsilon$ (omitted for brevity).
Treating the observed data set $\mathcal{D}$ as a single batch, 
AE estimate $\hat\phi$ using the same least-squared TD formula (\ref{eq:lstd}) of section 9.8 of \citet{BartoSutton:2020}.
Using $\hat\phi$, the estimated $R_{\hat\phi}(s,a)=\vec{\rho}(s,a)^\top\hat\phi$ that can be used as the covariate 
in the CCP estimator of  $\theta$ in (\ref{eq:ccpestimator}), and the other ``correction term''  $\hat Q_\varepsilon(s,a)$
can be estimated in similar fashion. The asymptotic distribution of this two-stage TD/CCP estimator of $\theta$  is 
 affected by ``estimation noise'' in the $\hat R$ and $\hat Q_{\varepsilon}$ functions
that can introduce bias and noise. To deal with this, AE propose a third estimation stage that uses  
a modified  score (i.e. gradient) of the  partial log-likelihood (\ref{eq:ccplf}) to create 
a robust (Neyman orthogonal) version of the score function \citep{lrspe:2022},
obtaining an improved estimator $\hat\theta$ robust to first stage estimation
noise in $\hat R$ and $\hat Q_\varepsilon$ by setting the modified score to zero.\footnote{\footnotesize
See \citet{khwaja:2025} for an alternative approach to AE's TD methods that they call RLTD-CCS that combines the TD algorithm 
from RL with the Conditional Choice Simulator (CCS) estimator of \citet{hmss:1994} that relies on forward simulations
of paths to produce a Monte Carlo estimate of the $Q_\theta(s,a)$  and hence is a model-based approach to estimation.}

AE also propose iterative estimators of $\theta$ for specifications where $\theta$ enters the reward function
in a non-linear-in-parameter fashion, i.e. $r_\theta(s,a)$,  using the ``target network'' idea
underlying the FVI algorithm (\ref{eq:dqn_target}) which they call the {\it approximate value iteration\/} (AVI)
estimator, and note that AVI can also overcome a ``limitation of the linear semi-gradient method is that it requires one to choose a series basis and also does not allow for high-dimensional state spaces'' \citep[p.~13]{AE:2022}.  However
 ``Compared to the semi-gradient approach, AVI is computationally more expensive as it requires solving $J$ prediction problems'' \citep[p.~13]{AE:2022}. They note that because the TD estimator is model-free, it will be
less efficient than full information maximum likelihood estimator (e.g. NFXP or NPL) when $p$ has a known parametric
functional form. On the other hand, ``if we make
the model for the transition density richer, while still keeping it parametric, the performance
of NFXP and NPL will start to degrade and become worse than TD estimation'' \citep[p.~24]{AE:2022}. They conjecture
that the TD estimator attains ``the semi-parametric efficiency bound when the transition density is unknown'' \citep[p.~24]{AE:2022}.

The GLADIUS estimator \citep{kang:2025} (Gradient-based Learning with Ascent-Descent for Inverse Utility learning from Samples) is model-free, does not require first-stage CCP estimation, and uses flexible function approximation. \citet{kang:2025} show that occupancy-matching methods (GAIL, IQ-Learn) minimize average Bellman error only on the expert's support, so $Q$ is unidentified off-support. GLADIUS 
addresses this by following the approach of \citet{antos:2008} and
 separately parameterizing $Q_{\phi}$ and $EV_{\psi}$ using two neural networks with separate
parameters $\phi$ and $\psi$, recovering the reward as $r(s,a) = Q_{\hat\phi}(s,a) - \beta EV_{\hat\psi}(s,a)$ via the soft Bellman equation (\ref{eq:smoothQ}).
GLADIUS maximizes a penalized likelihood where the penalty targets the Bellman error 
using formula (\ref{eq:BETDdiff}) to write the Bellman error as $\mbox{BE}(\phi,\psi)=\mbox{TD}(\phi)-\beta^2\mbox{VQ}(\phi,\psi)$
and estimate the parameters as a maxmin problem to find $(\hat\phi,\hat\psi)$, 
where the outer maximization searches for $\hat\phi$ to maximize the penalized likelihood, and
the inner minimization finds $\hat\psi$ to estimate $EV_\psi$ needed to compute 
the conditional variance term to correctly calculate the mean squared Bellman
error $\mbox{BE}(\phi,\psi)$ without requiring knowledge of $p(s'|s,a)$. Like the CCP estimator, GLADIUS imposes a normalizing action assumption for identification, using the Hotz-Miller inversion (\ref{eq:ccpinverse}) to recover remaining $Q$ values. Under 
realizability assumptions (i.e. that $Q=Q_{\phi^*}$ for some $\phi^*$ and
$EV=EV_{\psi^*}$ for some $\psi^*$), GLADIUS achieves global convergence with error $O(1/T) + O(1/N)$, the first such 
guarantee for methods minimizing mean squared Bellman residuals.

\subsection{Alternative Approaches}

Closely related to IRL is the Inverse Optimal Control (IOC) tradition in control theory. Originating with \citet{kalman:1964} for linear-quadratic systems, IOC was later adapted to complex, nonlinear biological behaviors by \citet{locomotion:2010} and \citet{berret:2011}. Their approach employed numerical bilevel optimization that, analogous to the nested fixed point algorithm in DDC, requires solving the full forward control problem within the inner loop of parameter estimation. The computational intractability of this inner loop for high-dimensional continuous domains led to the local probabilistic approximations of \citet{levine:2012}, which in turn enabled the deep, sampling-based Guided Cost Learning of \citet{guidedcostlearning:2016}, which merged the IOC tradition with IRL.

A parallel stream in cognitive science connects ``mentalizing'' to inverse reinforcement learning. \citet{baker:2009} modeled action understanding as inverse planning, a framework explicitly connected to the IRL literature by \citet{jaraettinger:2019}: observers infer unobservable mental states by inverting a forward RL model. These models extend standard IRL by recovering subjective epistemic and motivational states. The Bayesian Theory of Mind (BToM) framework treats agents as POMDP planners acting on potentially false beliefs \citep{baker:2017}, while the ``naive utility calculus'' \citep{jaraettinger:2016} grounds social reasoning in agent-specific cost-reward trade-offs. Although recent approaches like ToMnet \citep{rabinowitz:2018} scale via meta-learning without recovering explicit reward functions, this literature provides ways to model beliefs as separate from states and offers behavioral microfoundations for IRL.

A new direction of research integrates behavioral economics constraints directly into IRL. \citet{hoiles:2020}, for example, model agents under Rational Inattention, where acquiring information about the state $s$ incurs an information-theoretic cost. The agent first chooses an attention policy $\alpha(s'|s)$ (mapping true states $s$ to perceived signals $s'$) to maximize expected utility net of information costs. The observed choice probability $\pi(a|s)$ emerges as a mixture over these signals. The IRL problem is then to test whether observed choices are consistent with a reward function $r_\theta(s,a)$ solving this attention-constrained optimization. This revealed preference approach allows simultaneous recovery of the agent's reward parameters $\theta$ and their implicit information acquisition cost, providing a framework to test for and model bounded rationality directly from choice data.

\subsection{Applications}

We have surveyed IRL methods most relevant to DDC. Model-free methods avoid specifying $p(s'|s,a)$ but implicitly assume rational expectations. For counterfactual analysis involving changed dynamics, model-based methods remain necessary, since inaccurate models from finite data with limited coverage can compound errors in estimated rewards \citep{zeng:2024}.\footnote{\footnotesize \citet{schlaginhaufen:2024} show that with finite data, rank conditions for transfer \citep{rolland:2022} are insufficient; principal angles between training environments' transition laws bound reward error.} As \citet[p.~2]{gleave:2022} notes, ``algorithms based on MCE IRL have scaled to high-dimensional environments,'' with Maximum Entropy Deep IRL \citep{maxentdeepirl:2016} learning rewards from pixel observations, and GCL and AIRL scaling to continuous control tasks.

Large-scale applications include \citet{Barnes:2024} using IRL to infer route preferences from Google Maps data and \citet{zhao:2023} applying AIRL to taxi trajectories.\footnote{\footnotesize The Shanghai application succeeded because road networks provide deterministic transitions and the goal was matching observed trajectories rather than counterfactual analysis.} \citet{liang:2025} address the opacity of Deep IRL via knowledge distillation, training a surrogate multinomial logit (MNL) model on the Deep AIRL policy's soft labels to recover interpretable preference parameters. \citet{lee:2026} extend AIRL to serialized content consumption, using chapter text as state variables to study reader retention. Autonomous vehicles illustrate why IRL matters: \citet{waymo_bcsac:2023} combined imitation and RL (behavioral cloning--soft actor-critic, BC-SAC), achieving 38\% fewer safety failures than imitation alone, while \citet{remonda:2021} found RL agents with engineered lap-time rewards matched human speed but used fundamentally different control patterns, highlighting that engineered rewards fail to capture implicit human trade-offs.

\section{Discussion\label{sec:conclusion}}

Our survey reveals the broad similarity between the DDC literature which arose in the 1980s followed by IRL over a decade later.
Despite their different origins, they have a common goal: inferring a reward function $r$ presumed to govern the observed behavior of rational ``experts.''
Both fields use the same mathematical framework to model optimal dynamic decision making (e.g. MDPs based on ``soft Q'' and ``softmax'' formulas of section~\ref{sec:ddc}),
and both infer rewards using the same {\it bilevel optimization problem\/}: an ``outer'' maximization that searches over reward functions to maximize ``model fit''
(e.g. likelihood), and an ``inner'' minimization that solves the MDP for the optimal policy $\pi_r$.
However, the literatures differ on the method used to solve MDPs (DDC using deterministic algorithms
such as value and policy iteration, IRL using the stochastic iterative methods developed in RL), 
the interpretation of apparent randomness in individual behavior
 (DDC explaining randomness via unobserved state variables versus IRL's use of mixed strategies and entropy) and
on their primary motivation. DDC focuses on accurate inference for counterfactual policy evaluation, while IRL prioritizes automating complex behaviors 
like driving without explicit programming.

We discussed two key challenges facing both literatures: 1) the identification problem, and 2) the curse of dimensionality. 
The latter is problematic for DDC and IRL due to the need to repeatedly solve an MDP in the inside loop of the bilevel optimization problem. We discussed
the use of neural networks and randomization as two  ways to  circumvent the curse of dimensionality. While
there is no formal proof that they actually ``break'' this curse, the
astounding successes of deep reinforcement learning in RL applications such as Alpha-Zero's ability to beat any human player in chess or Go suggest  that these 
methods combined with sufficient computer power will enable us to use RL, DDC and IRL to tackle an increasing number of large scale, realistic problems. 
At the same time, these RL success stories call into question the assumption of human rationality: if Alpha-Zero beats all humans, 
we evidently cannot be perfectly rational optimizers. Instead humans may be better modeled as {\it boundedly rational\/} \citep{Simon:1957}, but 
there is relatively little work on how to adapt DDC and IRL to infer rewards of boundedly rational agents and understand whether suboptimal ``satisficing'' behavior
is due to inability to optimize or irrational beliefs.\footnote{\footnotesize \citet{tennis:2025} argue that
suboptimal play of professional tennis players is due to distorted subjective beliefs, and \citet{whoismorebayesian:2025} show that suboptimal behavior of
human subjects in lab experiments is due to subjective beliefs that differ from rational beliefs implied by Bayes Rule.}

To overcome the identification problem, we have to be willing
to impose strong {\it a priori\/} assumptions about rewards and beliefs, otherwise it is not possible to uniquely invert (identify) rewards and beliefs
consistent with observed behavior and make valid counterfactual predictions. But this means that counterfactual predictions are 
dependent on hard-to-verify assumptions. This is not just a problem for the credibility of counterfactual forecasts of economic
policies from DDC models: it also can limit the ability of IRL-trained agents to perform intelligently in 
environments that differ from the one they were trained in.
For example, if we identify rewards for an autonomous vehicle (AV) using data in an urban environment with transition probability $p(s'|s,a)$, but fail to disentangle preferences from dynamics,
the AV may perform well in the city but fail in a rural environment where traffic is  governed by a different transition probability $\lambda(s'|s,a)$.
Thus, mis-identified structural models may be subject to the same
problems  for counterfactual prediction that we noted for reduced-form models in the DDC literature
or for imitation learning methods such as {\it behavioral cloning\/} in the machine learning literature: they can fail to correctly predict
how humans change their behavior and rapidly adapt to new environments. 

Despite these challenges, when correct (or at least approximately correct) identifying assumptions are 
imposed, DDC and IRL models can make accurate counterfactual/out-of-sample predictions. For example, \citet{toddwolpin:2006} and \citet{lee:2026} demonstrated that a
structural model estimated using data on a control group accurately predicts the change in behavior
of a treatment group in field experiments, and \citet{zhuang:2024} showed that  AVs governed by reward functions trained by IRL
``can guide the direction of the policy network well in driving scenarios it had not encountered during training, thus validating the 
generalization capabilities of the reward functions across different drivers.'' (p. 8100).

We distinguished between  model-free vs model-based approaches to DDC, RL and IRL.
Most of the work in DDC is model-based, i.e. it requires the researcher to specify and estimate a model for the
transition probability $p(s'|s,a)$ which is estimated along with the reward function. While it is possible to estimate $p$ non-parametrically
in a first stage estimation before estimating $\{\beta,r\}$, this 
still entails the strong assumption of rational expectations and entails the burden of numerical integration with respect
to $p$ to solve the associated MDP for the optimal policy $\pi_r$. Model-free IRL methods have been developed that mimic the model-free
approaches developed in the RL literature such as Q-learning discussed in section~\ref{section:rl}, but instead of being done
in real-time in {\it online training\/} using the ``environment'' to simulate draws $s' \sim p(s'|s,a)$, model-free IRL can be done {\it offline\/} 
using historical data on transitions to fit parametric approximations (e.g. deep Q networks) $Q_\phi$, bypassing the need to estimate
$p$ or do explicit numerical integration.

Thus, model-free IRL has computational advantages and avoids the need to specify and explicitly estimate $p$. Yet,
model-free IRL is not assumption-free: it requires
the strong implicit assumption that individuals have rational expectations, which may be dubious as we noted above.
Confidence in model-free estimates of rewards requires sufficient data on $(s',s,a)$ transitions. Limited/sparse 
data can lead to poor estimates of $Q$ and noisy, unreliable estimates of the reward function, resulting in poor performance
similar to what one observes in RL after only a limited amount of initial training. Good performance from RL usually
requires substantial level of training, and this is why most of the successful applications of RL to solve complex
large scale problems are model-based. Even  in cases where we have sufficient data, such as the case of AVs
where  ``easily captured large-scale datasets of human driving to be used to train deep learning approaches to near human standard'' 
\citep[p.~14128]{mero:2022}, substantial additional offline training is often required to provide sufficient evidence that the  AV can be trusted
to operate safely in an online environment in real-time.
The success of offline training using computer simulations presumes we have an accurate model that covers a wide range of driving environments.
It should be obvious that even with an accurately estimated reward function,
extensive RL training with a poor model of the environment will not necessarily result in good performance. This leads to the conundrum: how
to develop an accurate model of the environment using limited data? 

This is a deep, unresolved question, and an example of where methods developed in RL have inspired new insights into human intelligence
in neuroscience and psychology, particularly with respect to the puzzle of how humans are able to adapt and learn to perform reasonably well on 
a wide range of novel tasks with only limited prior training and experience. \citet{theorybasedRL:2023} introduced {\it theory based reinforcement learning\/}
which ``posits that the agent learns the theory from experience using probabilistic inference and uses it together with an internal simulator 
to predict and evaluate the outcomes of different action sequences generated by an internal planner. It has captured patterns of human learning, exploration,
 and generalization in complex domains where model-free and simpler model-based RL approaches fail or learn rather differently.'' (p. 1331)

\clearpage
\begin{singlespace}
\bibliographystyle{aer}
\bibliography{our-cites.bib}
\end{singlespace}

\end{document}